\input harvmac
%
%\input epsf
%\draftmode
%
\noblackbox
%%%%%%%%%%%%%%%%%%%%%%%%%%%%%%%%%%%%%%%%%%%%%%%%%%%%%%%%%%%%%%%%%%%%%%%%%
\def\npb#1#2#3{{\it Nucl.\ Phys.} {\bf B#1} (19#2) #3}
\def\plb#1#2#3{{\it Phys.\ Lett.} {\bf B#1} (19#2) #3}
\def\prl#1#2#3{{\it Phys.\ Rev.\ Lett.} {\bf #1} (19#2) #3}

\def\atmp#1#2#3{{\it Adv.\ Theor.\ Math.\ Phys.} {\bf #1} (19#2) #3}
\def\jhep#1#2#3{{\it JHEP\/} {\bf #1} (19#2) #3}
%%%%%%%%%%%%%%%%%%%%%%%%%%%%%%%%%%%%%%%%%%%%%%%%%%%%%%%%%%%%%%%%%%%%%%%%%
% some stuff needed for figures:
%%%%%%%%%%%%%%%%%%%%%%%%%%%%%%%%%%%%%%%%%%%%%%%%%%%%%%%%%%%%%%%%%%%%%%%%%
\newcount\figno
\figno=0
\def\fig#1#2#3{
\par\begingroup\parindent=0pt\leftskip=1cm\rightskip=1cm\parindent=0pt
\baselineskip=11pt
\global\advance\figno by 1
\midinsert
\epsfxsize=#3
\centerline{\epsfbox{#2}}
\vskip 12pt
{\bf Fig.\ \the\figno: } #1\par
\endinsert\endgroup\par
}
\def\figlabel#1{\xdef#1{\the\figno}}
\def\encadremath#1{\vbox{\hrule\hbox{\vrule\kern8pt\vbox{\kern8pt
\hbox{$\displaystyle #1$}\kern8pt}
\kern8pt\vrule}\hrule}}
%%%%%%%%%%%%%%%%%%%%%%%%%%%%%%%%%%%%%%%%%%%%%%%%%%%%%%%%%%%%%%%%%%%%%%%%%%%

\def\frac#1#2{{#1 \over #2}}

\def\p{\partial}
\def\semi{\subset\kern-1em\times\;}
\def\bar#1{\overline{#1}}

\def\p{\partial}

\def\th{\theta}

\def\ad{\bar a}

\def\at{\tilde{a}}
\font\zfont = cmss10 %scaled\magstep1 
\font\zfonteight = cmss8 %scaled\magstep1 
\def\ZZ{\hbox{\zfont Z\kern-.4emZ}}
\def\ZZs{\hbox{\zfonteight Z\kern-.4emZ}}
%

%%%%%%%%%%%%%%%%%%%%%%%%%%%%%%%%%%%%%%%%%%%%%%%%%%%%%%%%%%%%%%%%%%%%%%%%%%%
\Title{\vbox{\baselineskip12pt
\hbox{hep-th/0012198}
\hbox{EFI-00-52}
\vskip-.5in}}
{\vbox{\centerline{Boundary String Field Theory of the  $D{\bar D}$ System}}}
%\bigskip
%\centerline{in String Theory}}}
\medskip\bigskip
\centerline{Per Kraus and Finn Larsen}
\bigskip\medskip
\centerline{\it Enrico Fermi Institute,} 
\centerline{\it  University of Chicago,} 
\centerline{\it  5640 S. Ellis Ave., Chicago, IL 60637, USA}
%\centerline{\tt pkraus, flarsen@theory.uchicago.edu}
\medskip
\baselineskip18pt
\medskip\bigskip\medskip\bigskip\medskip
\baselineskip16pt
\noindent
We develop the boundary string field theory approach to tachyon 
condensation on the $D{\bar D}$ system. Particular attention is paid to
the gauge fields, which combine with the tachyons in a natural
way. We derive the RR-couplings of the system and express
the result in terms of Quillen's superconnection. The result
is related to an index theorem, and is thus shown to be exact.

\Date{December, 2000}
%%%%%%%%%%%%%%%%%%%%%%%%%%%%%%%%%%%%%%%%%%%%%%%%%%%%%%%%%%%%%%%%%%%%%%%%%%%
%\cite{Takayanagi:2000rz}
\lref\Tak{
T.~Takayanagi, S.~Terashima and T.~Uesugi,
``Brane-antibrane action from boundary string field theory,''
hep-th/0012210.}
%%CITATION = HEP-TH 0012210;%%
%\cite{David:2000yn}
\lref\jdav{
J.~R.~David,
``Tachyon condensation using the disc partition function,''
hep-th/0012089.}
%%CITATION = HEP-TH 0012089;%%
%\cite{David:2000uv}:
\lref\hori{K.~Hori,``Linear Models of Supersymmetric D-Branes'',
[hep-th/0012179].}
\lref\david{
J.~R.~David,
``U(1) gauge invariance from open string field theory'',
JHEP {\bf 0010}, (2000) 017,
[hep-th/0005085].}
%%CITATION = HEP-TH 0005085;%%
%\lref\wittenscft{
%E.~Witten,
%``Noncommutative tachyons and string field theory,''
%hep-th/0006071.}
%%CITATION = HEP-TH 0006071;%%
\lref\tatar{R. Tatar, ``A Note on Non-Commutative Field Theory and Stability of
Brane-Antibrane Systems,'' hep-th/0009213.}      
%\cite{Dasgupta:2000ft}:
\lref\dmr{
K.~Dasgupta, S.~Mukhi and G.~Rajesh,
``Noncommutative tachyons,''
JHEP {\bf 0006}, (2000) 022,
[hep-th/0005006].}
\lref\schnabl{M. Schnabl, ``String field theory at large B-field and
noncommutative geometry,'' hep-th/0010034.}
\lref\ks{V .A. Kostelecky and S.Samuel, ``On a Nonperturbative Vacuum for the
Open Bosonic String,'' Nucl.Phys. {\bf B336} (1990) 263.}
\lref\sz{A. Sen and B. Zwiebach, ``Tachyon Condensation in String
Field Theory,'' hep-th/9912249.}
\lref\sendesc{A. Sen, ``Descent Relations Among Bosonic D-branes,'' Int.J. Mod.
Phys. {\bf A14} (1999) 4061, hep-th/9902105.}
\lref\sena{A.~Sen, ``Stable non-BPS bound states of BPS D-branes,''
\jhep{9808}{98}{010}, hep-th/9805019; 
``SO(32) spinors of type I and other solitons on brane-antibrane pair,''
\jhep{9809}{98}{023}, hep-th/9808141;
``Type I D-particle and its interactions,''
\jhep{9810}{98}{021}, hep-th/9809111; 
``Non-BPS states and branes in string theory,'' 
hep-th/9904207, and references therein.}
\lref\sennon{A. Sen, ``BPS D-branes on non-supersymmetric cycles,'' 
\jhep{9812}{021}{98}, [hep-th/9812031].}
\lref\bergman{O.~Bergman and M.~R.~Gaberdiel,
``Stable non-BPS D-particles,'' \plb{441}{98}{133}, hep-th/9806155.}
\lref\hhk{J. A. Harvey, P.Horava and P. Kraus, ``D-Sphalerons and the Topology
of String Configuration Space,'' hep-th/0001143.}
\lref\berk{N. Berkovits, ``The Tachyon Potential in Open Neveu-Schwarz String
Field Theory'', [hep-th/0001084].}
\lref\polchinski{J.~Polchinski, ``Dirichlet-Branes and Ramond-Ramond 
Charges,'' \prl{75}{95}{4724}, hep-th/9510017.}
\lref\senspinors{A.~Sen, ``$SO(32)$ Spinors of Type I and Other Solitons on 
Brane-Antibrane Pair,'' \jhep{9809}{98}{023}, hep-th/9808141.}
\lref\phk{P. Ho\v rava, ``Type IIA D-Branes, K-Theory, and Matrix Theory,'' 
\atmp{2}{99}{1373}, hep-th/9812135.}
\lref\yi{P. Yi, ``Membranes from Five-Branes and Fundamental Strings from
D$p$-Branes,'' \npb{550}{99}{214}; hep-th/9901159.}
\lref\senpuz{A. Sen, ``Supersymmetric World-volume Action for Non-BPS
D-branes,'' \hfill\break hep-th/9909062.}
\lref\taylor{W. Taylor, ``D-brane effective field theory from string
field theory,'' hep-th/0001201.}
\lref\cft{C. G. Callan, I. R. Klebanov, A. W. Ludwig and J.M Maldacena,
``Exact solution of a boundary conformal field theory,'' Nucl.Phys.
{\bf B422} (1994) 417, hep-th/9402113;
J. Polchinski and L. Thorlacius, ``Free fermion representation of a boundary
conformal field theory,''Phys. Rev. {\bf D50} (1994) 622, hep-th/9404008;
P.Fendley, H. Saleur and N. P. Warner, ``Exact solution of a massless
scalar field with a relevant boundary interaction,'' Nucl. Phys.
{\bf B430} (1994) 577, hep-th/9406125;
A. Recknagel and V. Schomerus, ``Boundary deformation theory and
moduli spaces of D-branes,'' Nucl. Phys. {\bf B545} (1999) 233,hep-th/9811237.}
\lref\rs{L.Randall and R. Sundrum, ``An alternative to compactification,''
Phys.Rev.Lett. {\bf 83} (1999) 4690, hep-th/9906064.}
\lref\agms{M.~Aganagic, R.~Gopakumar, S.~Minwalla and A.~Strominger,
``Unstable solitons in noncommutative gauge theory'',
hep-th/0009142.}
\lref\dbak{D.~Bak,
``Exact multi-vortex solutions in noncommutative Abelian-Higgs theory'',
hep-th/0008204.}
\lref\jmw{D.~P.~Jatkar, G.~Mandal and S.~R.~Wadia,
``Nielsen-Olesen vortices in noncommutative Abelian Higgs model'',
JHEP {\bf 0009} (2000) 018 , hep-th/0007078.}
\lref\gms{R.~Gopakumar, S.~Minwalla and A.~Strominger,
``Noncommutative solitons'',
JHEP {\bf 0005} (2000) 020 , hep-th/0003160.}
\lref\ci{L.~Cornalba,
``D-brane physics and noncommutative Yang-Mills theory'',
[hep-th/9909081]; N.~Ishibashi,
``A relation between commutative and noncommutative descriptions of  
D-branes'', [hep-th/9909176].}
%\cite{Witten:1981nf}
\lref\witind{
E.~Witten,
``Constraints On Supersymmetry Breaking,''
Nucl.\ Phys.\ {\bf B202}, 253 (1982).}
%%CITATION = NUPHA,B202,253;%%
\lref\index{
%\cite{Alvarez-Gaume:1984ig}
%\bibitem{Alvarez-Gaume:1984ig}
L.~Alvarez-Gaume and E.~Witten,
``Gravitational Anomalies,''
Nucl.\ Phys.\ {\bf B234}, 269 (1984);
%%CITATION = NUPHA,B234,269;%%
%\cite{Alvarez-Gaume:1983at}
%\bibitem{Alvarez-Gaume:1983at}
L.~Alvarez-Gaume,
``Supersymmetry And The Atiyah-Singer Index Theorem,''
Commun.\ Math.\ Phys.\ {\bf 90}, 161 (1983);
%%CITATION = CMPHA,90,161;%%
%\cite{Alvarez-Gaume:1983wp}
%\bibitem{Alvarez-Gaume:1983wp}
%L.~Alvarez-Gaume,
``A Note On The Atiyah-Singer Index Theorem,''
J.\ Phys.\ {\bf A16}, 4177 (1983);
%%CITATION = JPAGB,A16,4177;%%
%\cite{Friedan:1984xr}
%\bibitem{Friedan:1984xr}
D.~Friedan and P.~Windey,
``Supersymmetric Derivation Of The Atiyah-Singer Index And The Chiral Anomaly,''
Nucl.\ Phys.\ {\bf B235}, 395 (1984).}
%%CITATION = NUPHA,B235,395;%%

%\cite{Horava:1999jy}
\lref\petr{
P.~Horava,
``Type IIA D-branes, K-theory, and matrix theory,''
Adv.\ Theor.\ Math.\ Phys.\ {\bf 2}, (1999) 1373,
[hep-th/9812135].}
%%CITATION = HEP-TH 9812135;%%

%\cite{Wyllard:2000qe}
\lref\wyllard{
N.~Wyllard,
``Derivative corrections to D-brane actions with constant background  
fields'', [hep-th/0008125].}
%%CITATION = HEP-TH 0008125;%%

%\cite{Harvey:2000tv}
\lref\lumps{
J.~A.~Harvey and P.~Kraus,
``D-branes as unstable lumps in bosonic open string field theory,''
JHEP{\bf 0004}, (2000) 012,
[hep-th/0002117];
%%CITATION = HEP-TH 0002117;%%
%\cite{Berkovits:2000hf}
%\bibitem{Berkovits:2000hf}
N.~Berkovits, A.~Sen and B.~Zwiebach,
``Tachyon condensation in superstring field theory,''
Nucl.\ Phys.\ {\bf B587}, (2000) 147,
[hep-th/0002211];
%%CITATION = HEP-TH 0002211;%%
%\cite{deMelloKoch:2000ie}
%\bibitem{deMelloKoch:2000ie}
R.~de Mello Koch, A.~Jevicki, M.~Mihailescu and R.~Tatar,
``Lumps and p-branes in open string field theory,''
Phys.\ Lett.\ {\bf B482} (2000) 249,
[hep-th/0003031];
%%CITATION = HEP-TH 0003031;%%
%\cite{Moeller:2000jy}
%\bibitem{Moeller:2000jy}
N.~Moeller, A.~Sen and B.~Zwiebach,
``D-branes as tachyon lumps in string field theory,''
JHEP{\bf 0008}, (2000) 039,
[hep-th/0005036].}
%%CITATION = HEP-TH 0005036;%%

%\cite{Billo:1999tv}
\lref\ben{
M.~Billo, B.~Craps and F.~Roose,
``Ramond-Ramond couplings of non-BPS D-branes'',
JHEP{\bf 9906}, 033 (1999),
[hep-th/9905157].}
%%CITATION = HEP-TH 9905157;%%
%%CITATION = NUPHA,B188,513;%%
\lref\witncsft{E.~Witten,
``Noncommutative tachyons and string field theory'',
[hep-th/0006071].}
\lref\neksch{N.~Nekrasov and A.~Schwarz,
``Instantons on noncommutative $R^4$ and $(2,0)$ superconformal six  
dimensional theory'',
Commun.\ Math.\ Phys.\  {\bf 198} (1998) 689,
hep-th/9802068.}
\lref\harmor{J.~A.~Harvey and G.~Moore,
``Noncommutative tachyons and K-theory'',
hep-th/0009030.}
\lref\abs{M. F. Atiyah, R. Bott and A. Shapiro, ``Clifford Modules'',
Topology {\bf 3} suppl. 1 (1964) 3.}
\lref\absapp{E.~Witten,
``D-branes and K-theory'',
JHEP {\bf 9812} (1998) 019, [hep-th/9810188].}
%P.~Horava,
%``Type IIA D-branes, K-theory, and matrix theory'',
%Adv.\ Theor.\ Math.\ Phys.\  {\bf 2} (1999) 1373,
%hep-th/9812135.}
\lref\hklm{J.~A.~Harvey, P.~Kraus, F.~Larsen and E.~J.~Martinec,
``D-branes and strings as non-commutative solitons'',
JHEP {\bf 0007} (2000) 042,
[hep-th/0005031].
}
\lref\ekawai{T. Eguchi and H. Kawai, ``Reduction of Dynamical Degrees
of Freedom in the Large N Gauge Theory,'' Phys. Rev. Lett. {\bf 48}
(1982) 1063.}
\lref\senunique{A. Sen, ``Uniqueness of Tachyonic Solitons,''
hep-th/0009090.}
%\lref\senissue{A. Sen, ``Some Issues in Non-commutative Tachyon
%Condensation,'' hep-th/0009038.}
%\lref\gmsII{R. Gopakumar, S.Minwalla and A. Strominger,
%``Symmetry Restoration and Tachyon Condensation in Open String Theory,''
%hep-th/0007226.}
\lref\somerefs{
A.~Sen and B.~Zwiebach,
``Large marginal deformations in string field theory,''
JHEP {\bf 0010}, 009 (2000)
[hep-th/0007153];
W.~Taylor,
``Mass generation from tachyon condensation for vector fields on  D-branes,''
JHEP {\bf 0008}, 038 (2000)
[hep-th/0008033];
A.~Iqbal and A.~Naqvi,
``On marginal deformations in superstring field theory,''
hep-th/0008127.
}
\lref\soch{C. Sochichiu, ``Noncommutative Tachyonic Solitons. Interaction
with Gauge Field,'' hep-th/0007217.}
\lref\wittenstrings{E. Witten, ``Overview of K Theory Applied to Strings,''
hep-th/0007175.}
\lref\nati{N. Seiberg, ``A Note on Background Independence in Noncommutative
Gauge Theories, Matrix Model and Tachyon Condensation,'' hep-th/0008013.}
\lref\pioline{B. Pioline and A. Schwarz, ``Morita equivalence and T-duality
(or B versus Theta),''
JHEP {\bf 9908} (1999) 021;hep-th/9908019.} 
\lref\truncate{A.~Sen and B.~Zwiebach,
``Tachyon condensation in string field theory'',
JHEP {\bf 0003} (2000) 002,
hep-th/9912249;
N.~Moeller and W.~Taylor,
``Level truncation and the tachyon in open bosonic string field theory'',
Nucl.\ Phys.\  {\bf B583} (2000) 105, hep-th/0002237.}
\lref\tsolrefs{N.~Berkovits, A.~Sen and B.~Zwiebach,
``Tachyon condensation in superstring field theory'',
hep-th/0002211;
J.~A.~Harvey and P.~Kraus,
``D-branes as unstable lumps in bosonic open string field theory'',
JHEP {\bf 0004} (2000) 012 ,
hep-th/0002117;
R.~de Mello Koch, A.~Jevicki, M.~Mihailescu and R.~Tatar,
``Lumps and p-branes in open string field theory'',
Phys.\ Lett.\  {\bf B482}, 249 (2000), hep-th/0003031;
N.~Moeller, A.~Sen and B.~Zwiebach,
``D-branes as tachyon lumps in string field theory'',
JHEP {\bf 0008}, 039 (2000),
hep-th/0005036.}
\lref\gmsii{R.~Gopakumar, S.~Minwalla and A.~Strominger,
``Symmetry restoration and tachyon condensation in open string theory'',
hep-th/0007226.}
\lref\senvac{A.~Sen,
``Some issues in non-commutative tachyon condensation'',
hep-th/0009038.}
%\cite{Seiberg:1999vs}:
\lref\sw{
N.~Seiberg and E.~Witten,
``String theory and noncommutative geometry,''
JHEP {\bf 9909}, (1999) 032,
[hep-th/9908142].}
%%CITATION = HEP-TH 9908142;%%
\lref\bsft{D.~Kutasov, M.~Marino and G.~Moore,
``Some exact results on tachyon condensation in string field theory'',
hep-th/0009148.}
\lref\witcub{E.~Witten,
``Noncommutative Geometry And String Field Theory'',
Nucl.\ Phys.\  {\bf B268} (1986) 253.}
\lref\witbsft{E.~Witten,
``On background independent open string field theory'',
Phys.\ Rev.\  {\bf D46} (1992) 5467, hep-th/9208027.}
\lref\cds{A.~Connes, M.~Douglas and A.~Schwarz, ``Noncommutative Geometry
and Matrix Theory:Compactification on Tori,'' JHEP {\bf 02} (1998) 003;
hep-th/9711162.}
\lref\schom{V.~Schomerus, ``D-Branes and Deformation Quantization,''
JHEP {\bf 9906} (1999) 030; hep-th/9903205.}
\lref\countb{U. Venugopalkrishna, ``Fredholm operators associated with strongly
pseudo convex domains in $C^n$,'' J. Functional Anal. {\bf 9}
(1972) 349;      
 L. Boutet de Monvel, ``On the index of
Toeplitz operators of several complex variables,'' Inv. Math.
{\bf 50} (1979) 249.}
\lref\dh{M. R. Douglas and C. Hull, ``D-branes and the Noncommutative
Torus,'' JHEP {\bf 9802} (1998) 008;hep-th/9711165.}
\lref\furuuchi{K.~Furuuchi,
``Equivalence of projections as gauge equivalence on noncommutative  space'',
hep-th/0005199; K. Furuuchi, ``Topological Charge of $U(1)$ Instantons
on Noncommutative $R^4$'' hep-th/0010006.}
\lref\pho{P.~Ho, ``Twisted bundle on noncommutative space and U(1) instanton'',
hep-th/0003012.}
\lref\grnek{D. J. Gross and N.A.Nekrasov, ``Monopoles and Strings
in Noncommutative Gauge Theory,'' JHEP {\bf 0007} (2000) 034; hep-th/
0005204; D. J. Gross and N. A. Nekrasov, ``Dynamics of Strings in 
Noncommutative Gauge Theory,'' hep-th/0007204.}
\lref\nekrasov{N. A. Nekrasov, ``Noncommutative Instantons Revisited,''
hep-th/0010017.}
\lref\poly{A.~P.~Polychronakos,
``Flux tube solutions in noncommutative gauge theories,''
hep-th/0007043.}
%\lref\gorms{A.~ S.~Gorsky, Y.~M.~Makeeno and K.~G.~Selivanov,
%``On Noncommutative Vacua and Noncommutative Solitons,'' 
%hep-th/0007247.}
\lref\schom{
V.~Schomerus,
%``D-branes and deformation quantization,''
JHEP {\bf 9906}, 030 (1999)
[hep-th/9903205].}
\lref\hkm{J.~A.~Harvey, D.~Kutasov and E.~J.~Martinec,
``On the relevance of tachyons'',
[hep-th/0003101].}
\lref\senvanish{A.~Sen,
``Supersymmetric world-volume action for non-BPS D-branes,''
JHEP {\bf 9910}, 008 (1999)
[hep-th/9909062].}
\lref\dewit{B.~de Wit, J.~Hoppe and H.~Nicolai,
``On the quantum mechanics of supermembranes,''
Nucl.\ Phys.\  {\bf B305}, 545 (1988).}
\lref\coleman{See sec. 3.3 of S.Coleman, ``Quantum lumps and their classical
descendants,'' in {\it Aspects of Symmetry}, Cambridge University Press,
1985.}
\lref\corr{N.~Berkovits, A.~Sen and B.~Zwiebach,
``Tachyon condensation in superstring field theory'',
[hep-th/0002211];
P.~De Smet and J.~Raeymaekers,
``Level four approximation to the tachyon potential in superstring 
field  theory'', JHEP{\bf 0005} (2000) 051, [hep-th/0003220].
}
\lref\genact{
M.~R.~Garousi,
``Tachyon couplings on non-BPS D-branes and Dirac-Born-Infeld action'',
Nucl.\ Phys.\ {\bf B584} (2000) 284,
[hep-th/0003122];
E.~A.~Bergshoeff, M.~de Roo, T.~C.~de Wit, E.~Eyras and S.~Panda,
``T-duality and actions for non-BPS D-branes'',
JHEP{\bf 0005} (2000) 009,
[hep-th/0003221];
J.~Kluson,
``Proposal for non-BPS D-brane action'',
Phys.\ Rev.\ D {\bf 62} (2000) 126003,
[hep-th/0004106].}
\lref\hkl{J.~A.~Harvey, P.~Kraus and F.~Larsen,
``Exact noncommutative solitons'',
[hep-th/0010060].}

%%%%%%%%%%%%%%%%%%%%%%%%%%%%
%my refs
\lref\gms{R.~Gopakumar, S.~Minwalla and A.~Strominger,
``Noncommutative solitons'',
JHEP {\bf 0005} (2000) 020 , hep-th/0003160.}
\lref\bsrefs{M.~R.~Gaberdiel,
``Lectures on non-BPS Dirichlet branes'',
Class.\ Quant.\ Grav.\  {\bf 17} (2000) 3483,
[hep-th/0005029]; 
P.~Di Vecchia and A.~Liccardo,
``D branes in string theory. I'',
[hep-th/9912161].
}
\lref\acny{A.~Abouelsaood, C.~G.~Callan, C.~R.~Nappi and S.~A.~Yost,
``Open Strings In Background Gauge Fields'',
Nucl.\ Phys.\  {\bf B280} (1987) 599.}
\lref\birev{A.~A.~Tseytlin,
``Born-Infeld action, supersymmetry and string theory'',
[hep-th/9908105].}
\lref\ft{E.~S.~Fradkin and A.~A.~Tseytlin,
``Nonlinear Electrodynamics From Quantized Strings,''
Phys.\ Lett.\  {\bf B163} (1985) 123.}
\lref\at{O.~D.~Andreev and A.~A.~Tseytlin,
``Partition Function Representation For The Open Superstring Effective 
Action: Cancellation Of M\"{o}bius Infinities And Derivative Corrections To
Born-Infeld Lagrangian'',
Nucl.\ Phys.\  {\bf B311} (1988) 205.}
\lref\bsrrrefs{M.~Li,
``Boundary States of D-Branes and Dy-Strings'',
Nucl.\ Phys.\  {\bf B460} (1996) 351,
[hep-th/9510161];
M.~R.~Douglas,
``Branes within branes'',
[hep-th/9512077];
P.~Di Vecchia, M.~Frau, I.~Pesando, S.~Sciuto, A.~Lerda and R.~Russo,
``Classical p-branes from boundary state'',
Nucl.\ Phys.\  {\bf B507} (1997) 259,
[hep-th/9707068].}
%%CITATION = HEP-TH 9512077;%%
\lref\ghm{M.~B.~Green, J.~A.~Harvey and G.~Moore,
``I-brane inflow and anomalous couplings on D-branes'',
Class.\ Quant.\ Grav.\  {\bf 14} (1997) 47,
[hep-th/9605033].}
\lref\kw{C.~Kennedy and A.~Wilkins,
``Ramond-Ramond couplings on brane-antibrane systems'',
Phys.\ Lett.\  {\bf B464} (1999) 206,
[hep-th/9905195].}

%%%%%%%%%%%%%%%%%%
%more refs

\lref\bsft{D.~Kutasov, M.~Marino and G.~Moore,
``Some exact results on tachyon condensation in string field theory'',
[hep-th/0009148];
``Remarks on tachyon condensation in superstring field theory'',
[hep-th/0010108].}
\lref\witcub{E.~Witten,
``Noncommutative Geometry And String Field Theory'',
Nucl.\ Phys.\  {\bf B268} (1986) 253.}
\lref\ks{V .A. Kostelecky and S.Samuel, ``On a Nonperturbative Vacuum for the
Open Bosonic String,'' Nucl.Phys. {\bf B336} (1990) 263.}
\lref\sz{A.~Sen and B.~Zwiebach,
``Tachyon condensation in string field theory'',
JHEP {\bf 0003} (2000) 002,
[hep-th/9912249].}
\lref\witbsft{E.~Witten,
``On background independent open string field theory'',
Phys.\ Rev.\  {\bf D46} (1992) 5467, [hep-th/9208027].}
\lref\somcom{E.~Witten,
``Some computations in background independent off-shell string theory'',
Phys.\ Rev.\  {\bf D47} (1993) 3405,
[hep-th/9210065].}
\lref\liwit{K.~Li and E.~Witten,
``Role of short distance behavior in off-shell open string field theory'',
Phys.\ Rev.\  {\bf D48} (1993) 853,
[hep-th/9303067].}
\lref\shbsft{S.~L.~Shatashvili,
``Comment on the background independent open string theory'',
Phys.\ Lett.\  {\bf B311} (1993) 83,
[hep-th/9303143];
``On the problems with background independence in string theory'',
[hep-th/9311177].}
%\cite{Sen:1998ex}
\lref\FL{
A.~Sen,
``BPS D-branes on non-supersymmetric cycles,''
JHEP{\bf 9812}, 021 (1998)
[hep-th/9812031].}
%%CITATION = HEP-TH 9812031;%%
\lref\gsbsft{A.~A.~Gerasimov and S.~L.~Shatashvili,
``On exact tachyon potential in open string field theory'',
[hep-th/0009103].}
\lref\cubrefs{N.~Moeller and W.~Taylor,
``Level truncation and the tachyon in open bosonic string field theory'',
Nucl.\ Phys.\  {\bf B583} (2000) 105, hep-th/0002237. What else? 
Solitonrefs? Various sen papers}
\lref\tsolrefs{J.~A.~Harvey and P.~Kraus,
``D-branes as unstable lumps in bosonic open string field theory'',
JHEP {\bf 0004} (2000) 012 ,
[hep-th/0002117];
N.~Berkovits, A.~Sen and B.~Zwiebach,
``Tachyon condensation in superstring field theory'',
[hep-th/0002211];
R.~de Mello Koch, A.~Jevicki, M.~Mihailescu and R.~Tatar,
``Lumps and p-branes in open string field theory'',
Phys.\ Lett.\  {\bf B482}, 249 (2000), [hep-th/0003031];
N.~Moeller, A.~Sen and B.~Zwiebach,
``D-branes as tachyon lumps in string field theory'',
JHEP {\bf 0008}, 039 (2000),
[hep-th/0005036].}
\lref\gms{R.~Gopakumar, S.~Minwalla and A.~Strominger,
``Noncommutative solitons'',
JHEP {\bf 0005} (2000) 020 , hep-th/0003160.}
\lref\cori{L.~Cornalba,
``D-brane physics and noncommutative Yang-Mills theory'',
[hep-th/9909081]; N.~Ishibashi,
``A relation between commutative and noncommutative descriptions of  
D-branes'',[hep-th/9909176].}
\lref\tsebsft{A.~A.~Tseytlin,
``Sigma model approach to string theory effective actions with tachyons'',
[hep-th/0011033].}
\lref\anbsft{O.~Andreev,
``Some computations of partition functions and tachyon potentials 
in  background independent off-shell string theory'',
[hep-th/0010218].}

\lref\ghm{M.~B.~Green, J.~A.~Harvey and G.~Moore,
``I-brane inflow and anomalous couplings on D-branes'',
Class.\ Quant.\ Grav.\  {\bf 14} (1997) 47
[hep-th/9605033].}
\lref\kw{C.~Kennedy and A.~Wilkins,
``Ramond-Ramond couplings on brane-antibrane systems'',
Phys.\ Lett.\  {\bf B464} (1999) 206,
[hep-th/9905195].}
\lref\grefs{I.~Affleck and A.~W.~Ludwig,
``Universal noninteger 'ground state degeneracy' in critical quantum 
systems'', 
Phys.\ Rev.\ Lett.\  {\bf 67} (1991) 161.}
\lref\tendim{J.~A.~Harvey, S.~Kachru, G.~Moore and E.~Silverstein,
``Tension is dimension'',
JHEP {\bf 0003} (2000) 001,
[hep-th/9909072].}
\lref\marcs{N.~Marcus and A.~Sagnotti,
``Group Theory From 'Quarks' At The Ends Of Strings'',
Phys.\ Lett.\  {\bf B188} (1987) 58.}
\lref\marc{N.~Marcus,
``Open String And Superstring Sigma Models With Boundary Fermions'',
DOE-ER-40423-09-P8 (1988).}
\lref\lwfer{F.~Larsen and F.~Wilczek,
``Geometric entropy, wave functionals, and fermions'',
Annals Phys.\  {\bf 243} (1995) 280, [hep-th/9408089].}
\lref\sennor{D.~Ghoshal and A.~Sen,
``Normalisation of the background independent open string field theory  
action'', JHEP {\bf 0011} (2000) 021,
[hep-th/0009191].}
\lref\senrev{A.~Sen,
``Non-BPS states and branes in string theory'',
[hep-th/9904207].}
\lref\senuniv{A.~Sen,
``Universality of the tachyon potential'',
JHEP {\bf 9912} (1999) 027, [hep-th/9911116].}
\lref\quillen{D.~Quillen, ``Superconnections and the Chern character'',
Topology {\bf 24} 89-95 (1985).}
\lref\book{N. Berline, E. Getzler, and M. Vergne, Heat kernels and Dirac
operators'', Springer-Verlag (1991).}

\newsec{Introduction}

Tachyon condensation is the conceptually simple process of
 fields rolling down a potential towards a  minimum. 
The technical challenge  of describing this phenomenon in string theory 
is that such a shift of the vacuum cannot be studied in the standard
first quantized formalism; the physics is 
necessarily off-shell and so in the domain of string field theory (SFT).
%
% obstacle in describing this process in string theory 
%is that constant tachyon fields do not satisfy the equations of motion, 
%because the tachyon ``mass'' is nonvanishing. The question is therefore 
%necessarily off-shell and in the domain of string field theory (SFT). 
The purpose of this article is to develop the boundary SFT 
for the basic unstable system of Type II string theory,
%for one of the simplest systems with a tachyonic instability, 
the $D\overline{D}$-system.  The physics of 
both  BPS D-branes and ``wrong $p$''
non-BPS D-branes can be recovered from this system by appropriate projections.
We will particularly emphasize the role 
of gauge fields and their interplay with the tachyons. We also derive 
the RR-couplings of the branes, expressing the results in a compact
form using Quillen's superconnection. 

The best developed approach to SFT is the cubic SFT for the open bosonic
string \witcub . 
This version gives a well defined theory with interactions that are
closely tied to worldsheet geometry.  
%
%is logically 
%satisfying and conceptually straightforward. 
Additionally, the truncation to 
the first few string levels provides a practical method for studying tachyon 
condensation, 
yielding quantitative results \refs{\ks,\sz}. 
Much intuition about the subject has been developed this way. 
The difficulty in using the cubic SFT is that an infinite number of component
fields acquire expectation values as the tachyon condenses.  
This  makes it difficult to obtain an analytical description of 
tachyon condensation.

In recent months the background independent version of open SFT 
\refs{\witbsft,\somcom,\shbsft} has been established 
as a viable alternative to the cubic SFT \refs{\bsft,\gsbsft}. 
Classical solutions in string theory are conformal field theories 
(CFTs) so it is natural to interpret SFT as a theory on the 
space of all two dimensional field theories, conformal or not. 
Background independent SFT is an attempt to make this concrete. 
It was originally derived through the Batalin-Vilkovisky formalism, 
but it is also possible to take a more intuitive approach, 
generalizing ordinary sigma models \refs{\ft,\acny,\at,\anbsft,\tsebsft}. 
This is the strategy we pursue.

The idea is the following. We are interested in open string field theory, 
so the closed strings are treated as an on-shell background.
For conformally invariant theories it is well known that the 
classical spacetime action of the open string theory is given by 
the partition  function on the disc. The new ingredient in string 
field theory is that we allow  boundary interactions which break 
conformal invariance  --- this is precisely what we mean by taking 
open strings off-shell. The working assumption is that the spacetime
action can be identified with the partion function also in this
more general setting. All of our computations thus boil down to computing 
partition functions on the disk of theories which are conformal in 
the bulk but not necessarily on the boundary.  

This approach to SFT can also be motivated
from a different point of view \hkm .
The renormalization group flow in the space of 2D quantum field theories, 
from one 2D CFT to another, is characterized by the $c$-functional. 
On surfaces with boundaries the flow between theories with identical 
conformal bulk but different boundary theories is similarly 
characterized by the $g$-functional \grefs , also known as the boundary 
entropy. The boundary entropy is a 
measure of the size of phase-space, as the name suggests, but it also 
measures D-brane tension \tendim . 
The tension is thus a function of the boundary interaction,
 suggesting an interpretation as an action on the space of 
theories, whether conformal or not. Since the $g$-functional is in fact
nothing other than the partition sum we return to the identification 
between the disc partition sum and the spacetime action. 

For the bosonic string, it is not quite correct to equate the spacetime
action with the disk partition sum, but the needed modifications can 
be obtained either from the BV formalism or from general 
considerations \shbsft .
However, for the  superstring  we are not
aware of any problems with this identification
(for discussion see \bsft ), and the fact that we obtain reasonable 
results serve {\it a posteriori} as  further justification.

Let us consider the strategy a little more concretely. The starting point is 
an on-shell closed string background described by one of the standard 
methods. It is convenient to describe it in the Schr\"{o}dinger 
representation as a closed string wave functional $\Psi_{\rm bulk}$. 
We consider very simple bulk states, 
corresponding to either the NS-NS or RR vacua,
but generalizations are possible. The remaining
ingredient is the boundary interaction, described by the
wave functional $\Psi_{\rm bndy}$.
The boundary wave functional is defined for general boundary interactions,
but those giving rise to free field theories are singled out as being
particularly simple.
 The final step is to combine the two 
ingredients by projecting the boundary wave functional on to the bulk wave
functional.  The result is the disk partition sum, which we then identify
with the spacetime action for open string fields, as explained above. 
In view of the significance of the boundary interaction, this 
form of SFT is sometimes referred to as boundary SFT  rather than background 
independent SFT; either way it is BSFT.

% on the open strings, described as
%a boundary interaction. The content of this theory is summarized in a 
%boundary wave functional $\Psi_{\rm boundary}$. The boundary wave functional 
%is defined for general theories but free theories play a special role 
%by making it particularly explicit. The final step is to combine the two 
%ingredients by projecting the boundary wave functional on to the bulk state 
%vector. The result can be interpreted as the generating functional of 
%open string 
%correlators or alternatively as the spacetime action, as explained above. 
%In view of the significance of the boundary interaction this 
%form of SFT is referred to as boundary SFT, rather than background 
%independent SFT; either way it is BSFT.

The contributions of the present paper fall in the following categories:
\item{(1)}
We compute the effective action for the tachyons of the $D\overline{D}$
system, and show that lower dimensional D-branes arise as solitons in
the expected fashion. All other D-branes -- non-BPS and BPS -- appear
as special cases of the $D{\bar D}$. Our organization of the 
computation is somewhat different from some recent presentations, and 
offers -- in our opinion --  conceptual and computational advantages.
As discussed above, our method is similar to the sigma model 
approach \refs{\ft,\acny,\at,\anbsft,\tsebsft,\jdav}.

\item{(2)} 
We include  gauge fields on the branes. The introduction of 
gauge fields {\it via} boundary fermions is explained in some detail.
It becomes apparent that in the $D\overline{D}$-system it is economical 
to consider gauge fields and tachyons simultaneously,
as the problem thus acquires its natural generality.
 As a concrete result we derive various
 terms in the combined tachyon-gauge field  action.
\item{(3)}
We consider the system in the background of a constant RR potential. 
The role of D-branes as sources of RR-charges makes this a natural 
problem. In the presence of the RR-background the modings of
fermions and bosons are identical, effectively reducing the problem 
to the fermion
zero-modes. The couplings we derive are summarized in the action
\eqn\coupl{
S = T_{D9}\int\! C\wedge {\rm Str} ~e^{2\pi\alpha^{\prime}i{\cal F}}~,
}
where ${\cal F}$ is the curvature of the superconnection \refs{\quillen,\book}
\eqn\supcon{
i{\cal A}= \left(
\matrix{
	iA^{+} & {\bar T}  \cr
	T & 	iA^{-}
}\right)~.}
This action generalizes the well-known RR-couplings of BPS D-branes 
to the $D\overline{D}$ system,
 and was conjectured by Kennedy and Wilkins \kw . 
Additionally, we find the corresponding couplings for the non-BPS
D-branes. From the result \coupl\ we 
verify that lower dimensional D-branes 
described as solitons carry the correct RR-charge. 
 We also discuss some of the  connections to index theorems.

\bigskip

The organization of this paper is as follows.
In section 2 we introduce the basic concepts of bulk and
boundary wave functionals and the tachyon effective action.
In section 3 we add non-abelian degrees of freedom 
 to the problem by discussing boundary 
fermions in some detail.  We also define the partition function for the 
$D\overline{D}$ system.  
In section 4 we discuss
lower dimensional branes as solutions in the resulting theory,
and  derive terms in the combined tachyon gauge field 
action.   In section 5
we consider the couplings to background RR potentials 
and discuss their implications.   We conclude with a discussion in 
section 6.  
 
%%%%%%%%%%%%%%%%%%%%%%%%%%%%%%%%%%%%%%%%%%%%%%
\newsec{Boundary String Field Theory}
The purpose of this section is to state the general procedure defining 
boundary string field theory. We also work out simple examples, yielding 
results needed later.

As explained in the introduction, we wish to compute the path 
integral over all fields on the unit disk in the presence of specified 
boundary interactions. It is convenient to perform the analysis 
in several steps:
\item{(1)}
Integrate over fields in the bulk. The result is a closed string wave 
functional,  a functional of the fields restricted to the boundary 
of the disk.
\item{(2)}
Include boundary interactions, as described by a
boundary wave functional.
\item{(3)} 
Project the boundary wave functional on to the bulk wave functional. 
The result is
the disc partition function, a functional of the spacetime fields appearing as 
 boundary couplings. 
\item{(4)}
The partition function thus computed is divergent and must be 
regularized and renormalized. We do so by zeta function methods. 
In superstring theory, the resulting renormalized partion function 
is identified with the spacetime action.
\bigskip
\noindent
In the following we make this procedure explicit through several 
important examples.

\subsec{\bf The Bosonic String Partition Function}
We begin by computing the disc partition function of the bosonic 
string, formally defined as
\eqn\b{Z = \int \! {\cal D}X \, e^{-(S_{{\rm bulk}}+ S_{{\rm bndy}})}~,}
with 
\eqn\c{S_{{\rm bulk}} = {1 \over 4\pi \alpha'}\int\! d^2x\, \sqrt{\gamma}
\gamma^{ab}\partial_a X^\mu \partial_b X_\mu~.}
As explained above, we first perform the integral over bulk field
configurations with fixed boundary conditions.
 We write the metric on the disc as
\eqn\a{ds^2 = d\rho^2 +\rho^2 d\tau^2~,}
with boundary at $\rho =1$, and specify the field on the boundary as
\eqn\d{X^\mu|_{\rho=1} = X_{0}^\mu+\sqrt{\alpha^{\prime}\over 2}
\sum_{n=1}^{\infty} \left(X_n^\mu e^{in\tau}+X_{-n}^\mu e^{-in\tau}\right)~.}
The unique regular solution to the bulk equation of motion, 
$\nabla^2 X^\mu=0$, is
\eqn\e{X^\mu = X_{0}^\mu+\sqrt{\alpha^{\prime}\over 2}\sum_{n=1}^{\infty}
\rho^n\left(X_n^\mu e^{in\tau} + X_{-n}^\mu e^{-in\tau}\right)~,}
and the bulk action evaluated on this solution is
\eqn\f{S_{{\rm bulk}} = {1 \over 4\pi \alpha'}\int \! 
d\tau X^\mu \partial_\rho
X^\mu |_{\rho=1} = {1 \over 2} \sum_{n=1}^{\infty} n X_{-n}^\mu X_{n}^\mu~.}
The corresponding bulk wave functional
\eqn\bulksv{
\Psi_{\rm bulk} = e^{-S_{\rm bulk}} = \exp\left(-
{1 \over 2} \sum_{n=1}^{\infty} n X_{-n}^\mu X_{n}^\mu~ \right)~,
}
characterizes the conformally invariant closed string vacuum.
The computation only took the saddle point contribution into account. 
This is justified because the overall factor from the fluctuations around the 
classical field is independent of the boundary field \d , 
and therefore irrelevant for our further considerations.

The result obtained so far is the starting point irrespective of the 
boundary interactions. 
We now consider the quadratic boundary interaction~\refs{\somcom,\bsft} 
\eqn\kyy{S_{{\rm bndy}} =  \int\! {d\tau \over 2\pi} \, u X^2 
 = u X_0^2 + u\alpha^{\prime} \sum_{n=1}^{\infty} X_{-n} X_{n}~,}
where the index on $X$ is omitted because we focus on a specific $X$.
This boundary interaction breaks conformal invariance and therefore takes 
the theory 
off-shell. It is a relevant interaction, 
inducing an RG flow between two CFTs. 
This flow describes tachyon condensation: the boundary interaction \kyy\ 
is interpreted 
in spacetime as a tachyon profile of the form $T(X)=uX^{2}$, as explained 
in \bsft .

The above evaluation of the bulk integral amounts to 
summing over all field configurations 
at $\rho<1$; what remains then is to integrate over the fields at $\rho=1$. 
The result is
\eqn\myy{Z(u) = \int \!{dX_0 \over \sqrt{2 \pi \alpha'}}
 \prod_{n=1}^\infty {dX_n dX_{-n}\over 4\pi}
e^{-(S_{{\rm bulk}} + S_{{\rm bndy}})}
= {1 \over \sqrt{2\alpha'u}} \prod_{n=1}^\infty \left(
{ 1 \over n +2\alpha'u}
\right)~.}
We chose a convenient measure; the factors in the denominator affect
only the $u$ independent normalization factor, which we are not 
keeping track of anyway. The infinite product is divergent (more accurately, 
$1/Z(u)$ diverges) and needs to be regularized. We use zeta function 
regularization:
\eqn\o{\eqalign{\prod_{n=1}^\infty \left( {1 \over n +2\alpha'u}\right)&
%= \exp{\left\{-\sum_{n=1}^\infty \ln(n +2\alpha'u)
%  \right\}} 
 = \exp{\left\{ {d \over ds} \sum_{n=1}^\infty 
\left(  n +2\alpha'u \right)^{-s}\right\}_{s=0}}
= \exp{\left\{ {d \over ds}  
\left[\zeta(s,2\alpha'u) - (2\alpha'u)^{-s}\right]\right\}_{s=0}} 
 \cr & \cr
 %= \exp{\left\{ {d \over ds}  
%\left[\zeta(s,2\alpha'u) - (2\alpha'u)^{-s}\right]\right\}_{s=0}}     
&= \exp{\left\{\ln \Gamma(2\alpha'u) - {1 \over 2}\ln 2\pi + \ln 2\alpha'u 
\right\}} 
%& \cr
= {2\alpha 'u ~\Gamma(2\alpha'u) \over \sqrt{2\pi}}~,
}}
where we used the zeta function 
\eqn\p{\zeta(z,q) = \sum_{n=0}^\infty {1 \over (q+n)^z}~.}
Our final result for the partition function is thus
\eqn\q{Z(u) = {\sqrt{2\alpha'u}~ \Gamma(2\alpha'u) \over  \sqrt{2 \pi}}~.}
This expression differs by a factor of $e^{2\alpha' u\gamma}/\sqrt{2\pi}$
from that of \somcom\ due to a different choice of regularization scheme. 
This difference does not affect any physical quantities such as 
D-brane tensions. 

\subsec{{\bf Including Gauge Fields}}
Another important example is that of an abelian gauge field, coupled to the 
worldsheet via the boundary term
\eqn\r{S_{A} =  -i \int\!  {d\tau} \, A_\mu(X^\mu)\dot{X}^\mu~, }
where $\dot{X} = dX/d\tau$. This coupling exhibits the gauge invariance
\eqn\s{\delta A_\mu  =\partial_\mu \alpha~.}
In the case of a constant field strength  we can write
\eqn\t{A_\mu = - {1 \over 2}F_{\mu\nu}X^\nu~,}
and the boundary coupling is
\eqn\u{S_{A} = {i \over 2} \int\! {d\tau}\, 
F_{\mu\nu}\dot{X}^\mu X^\nu~.}
Using mode expansions of the form \e\ gives
\eqn\v{S_{A} =  
\pi\alpha' F_{\mu\nu}\sum_{n=1}^\infty n\, X^\mu_{-n}
X_{n}^\nu~.}
So for constant field strengths and 
tachyon profile $T = \sum_\mu u_\mu (X^\mu)^2$,
and using the results \kyy\ and \myy , 
the total action is 
\eqn\w{S_{{\rm bulk}} + S_{{\rm bndy}} =  u_\mu (X_0^\mu)^2
+ {1 \over 2}\sum_{n=1}^\infty X^\mu_{-n} 
\left\{(n+2\alpha'u_\mu)\delta_{\mu\nu} +  2\pi \alpha' n F_{\mu\nu}
\right\} X_n^\nu~.}
(sum over $\mu, \nu$ implied).  The partition function is obtained from the
obvious higher dimension generalization of \myy .

To proceed we skew-diagonalize $F_{\mu\nu}$, with eigenvalues 
$f_{\beta}\equiv F_{2\beta,2\beta+1}$.  The integration gives
\eqn\x{Z(u,F) = 
%\left[\prod_\mu {1 \over \sqrt{2\alpha' u_\mu}}\right]
\int \! \prod_\mu {dX_0^\mu \over \sqrt{2\pi \alpha'}} 
e^{-u_\mu (X_0^\mu)^2}
\prod_{\beta=0}^{(d-1)/2} \prod_{n=1}^\infty  {1 \over
(n+2 \alpha' u_{2\beta})(n+2\alpha' u_{2\beta+1})
+(2\pi\alpha^{\prime} f_{\beta}n)^2}.}  
For vanishing tachyon, $u_{\mu}=0$, the zeta function prescription is
\eqn\y{ \prod_{n=1}^\infty {1 \over n^2 + a^2 n^2} = 
e^{2\zeta'(0)}(1+a^2)^{-\zeta(0)} = {\sqrt{1+a^2} \over 2\pi}~,}
which yields
\eqn\z{Z(0,F) =C\int\! d^{d+1}X_0\,\sqrt{\det 
\left[\delta_{\mu\nu}+ 2\pi \alpha' F_{\mu\nu}\right]}~,}
with the overall normalization constant working out to be 
$C= (4 \pi^2\alpha')^{-(d+1)/2}.$ 
This is the standard Born-Infeld action, as expected.

For tachyon profiles that are identical in pairs, $u_{2\beta}=u_{2\beta+1}$,
\x\ reduces to \anbsft\
\eqn\zmred{
Z(u,F) = \prod_{\beta=0}^{(d-1)/2} 
\sqrt{1+(2\pi\alpha^\prime f_\beta)^2}
~|Z({u_{2\beta}\over 1+2\pi\alpha^\prime if_\beta})|^2 ~,
}
where $Z$ is the partition function \q\ when a single $u_\mu$ is turned on.

To include non-abelian tachyons and gauge fields on multiple D-branes  
replace $T$ and $A_\mu$ by Hermitian matrices and introduce a path ordered 
exponential,
\eqn\aa{e^{-S_{{\rm bndy}}} 
= {\rm Tr}\,P e^{-\int\!{d\tau}\,
(T(X) +i A_\mu \dot{X}^\mu)}.}
Path ordering is necessary in order that the partition function be invariant
under the gauge transformation
\eqn\ab{\eqalign{\delta A_\mu &=
\partial_\mu \alpha + i[\alpha,A_\mu] \cr
\delta T &= i[\alpha,T].}}
Due to the nontrivial matrix orderings, no closed form expression is known
for the general non-abelian partition function.   Another approach to 
including non-abelian degrees of freedom is to introduce auxilliary boundary 
fermions instead of explicit Chan-Paton factors.  We will have more to say
about this approach when we come to  the superstring. 

\subsec{\bf Preliminaries for the Superstring}
We would like to carry out the corresponding computations for
the superstring. The starting point is the bulk action for the NSR string
\eqn\ac{S_{{\rm bulk}} ={1 \over 4\pi}\int\! d^2z\, \left(
{2 \over \alpha'}\partial X^\mu \bar{\partial}X_\mu 
+ \psi^\mu \bar{\partial} \psi_\mu
+\tilde{\psi}^\mu \partial \tilde{\psi}_\mu \right)~.}
We work in the NS sector so the fermions are anti-periodic on the disk. 
The mode expansion of $\psi^\mu$ at the boundary of the disc is
\eqn\af{\psi^\mu(\tau) =  
\sum_{r={1 \over 2}}^{\infty} \left( \psi^\mu_r e^{ir \tau} +  
\psi^\mu_{-r} e^{-ir \tau}\right)~.}
Fermions are simpler in rectangular coordinates than in polar so
it is convenient to extend into the bulk by going to the upper 
half-plane. We therefore temporarily complexify the coordinate
$\tau$, imposing regularity as ${\rm Im} \tau\to\infty$.
This prescription gives\foot{The left-moving classical field 
$\psi^\mu$ necessarily depends on both holomorphic and anti-holomorphic
coordinates; indeed, otherwise the on-shell action would vanish
because it is proportional to $\bar{\partial}\psi^\mu$. The treatment
of fermion wave functionals is discussed  in {\it e.g.}\lwfer .}
\eqn\aff{\psi^\mu(\tau) =  
\sum_{r={1 \over 2}}^{\infty} \left( \psi^\mu_re^{ir\tau} +  
\psi^\mu_{-r}e^{-ir{\bar\tau}}~\right)~,}
and the bulk action works out to be 
\eqn\ag{ S_{{\rm bulk}} = {1 \over 2} \sum_{n=1}^{\infty} n X_{-n}^\mu 
X_{n}^{\mu}+i\sum_{r={1 \over 2}}^{\infty} \psi_{-r}^\mu \psi_{r}^{\mu}~.}
The right-moving fermions are ${\tilde\psi}^{\mu}(\tau)=\psi^\mu(\tau)$, so 
they simply contributed a factor of $2$.

The next step is to introduce boundary interactions. An important 
principle  that we must respect is worldsheet supersymmetry. We do
so by working in boundary superspace with coordinates 
$\hat{\tau}=(\tau,\theta)$
and 
\eqn\ah{{\bf X}^\mu = X^\mu + \sqrt{\alpha'}\theta \psi^\mu~, \quad\quad
D= \partial_\theta + \theta \partial_\tau~.}
The simplest example of a boundary action corresponds to the gauge 
field. For an 
abelian gauge field we use the supersymmetric generalization of \r,
\eqn\ai{S_A = -i \int\! {d\tau} d\theta\, A_\mu ({\bf X})D{\bf X}^\mu
 = -i \int\! {d\tau} \left[ A_\mu(X)\dot{X}^\mu 
+ {1 \over 2}\alpha' F_{\mu\nu}\psi^\mu 
\psi^\nu \right]~.}
It is a simple matter to expand this boundary action in modes and compute the 
partition function. The unsurprising result \birev\ 
is that the fermion determinant
is independent of $F_{\mu\nu}$ so that we recover the 
Born-Infeld action \z .

\newsec{{\bf Boundary Fermions}}
In our approach to the $D{\bar D}$ system boundary interactions will be
introduced using auxiliary boundary fermions.
%We would like to study tachyon condensation in the $D{\bar D}$ system. 
%The appropriate boundary interactions are introduced using boundary 
%fermions. 
In order to motivate and explain the construction we first 
consider the related question of describing
non-abelian gauge fields. It will turn out that tachyons are similarly
described.

\subsec{\bf Non-abelian Gauge Fields}
Boundary interactions for non-abelian gauge fields in superstring theory must
simultaneously preserve spacetime gauge invariance and worldsheet 
supersymmetry. There are
several ways to achieve this.

One option is to use supersymmetric path ordering and consider
\eqn\aj{e^{-S_A} = {\rm Tr}\, \hat{P} 
e^{i \int\! {d\tau} d\theta\, A_\mu ({\bf X})D{\bf X}^\mu}~.}
The $\hat{P}$ symbol is
\eqn\ak{\hat{P}~ e^{\int \! d\hat{\tau}\, M(\hat{\tau})}
= \sum_{N=0}^\infty \int \! d\hat{\tau}_1 \ldots  d\hat{\tau}_N
\Theta(\hat{\tau}_{12})\Theta(\hat{\tau}_{23}) \ldots 
\Theta(\hat{\tau}_{N-1,N})
M(\hat{\tau}_1) \ldots M(\hat{\tau}_N)~,}
where $\hat{\tau}_{12} = \tau_1 -\tau_2 - \theta_1 \theta_2$, and $\Theta$ is
the step function.  The delta function term in the expansion
$\Theta(\hat{\tau}_{12}) = \Theta(\tau_1 -\tau_2) - 
\theta_1 \theta_2 \delta(\tau_1-\tau_2)$ gives contact terms which 
are essential 
for worldsheet supersymmetry.  An interesting feature is that these 
same contact 
terms are crucial for gauge invariance \at, 
as they contribute the $[A_\mu,A_\nu]$ 
in the non-abelian field strength $F_{\mu\nu}~$.  Indeed, performing 
the $d\theta$ 
integrals in \aj\ gives
\eqn\al{e^{-S_A} = {\rm Tr}~ P e^{i\int \! {d\tau} \, \left
[ A_\mu(X)\dot{X}^\mu + {\alpha' \over 2}F_{\mu\nu}\psi^\mu 
\psi^\nu \right]}~,}
with $F_{\mu\nu} = \partial_\mu A_\nu -\partial_\nu A_\mu- i[A_\mu,A_\nu]$.

The drawback of describing non-abelian interactions this way is that path 
ordering is awkward, whether supersymmetric or not. An alternative 
that is sometimes 
convenient is to trade the explicit path ordering in \aj\ for a path 
integral over boundary fermions $\eta^a$ in the fundamental representation
of the gauge group \at . 
Introducing the boundary superfield 
$\hat{\eta}^a = \eta^a + \theta \zeta^a$,
\aj\ can be rewritten as
\eqn\am{e^{-S_A} = \int\! D\hat{\eta}\, D\hat{\bar{\eta}} \,
e^{- \int\! 
{d\tau} d\theta\,\left[\hat{\bar{\eta}}^a D \hat{\eta}^a  
  - i \hat{\bar{\eta}}^a A_\mu^{ab}({\bf X})D{\bf X}^\mu \hat{\eta}_b 
  \right]}~.}

Rather than \aj, \al, or \am, we will use yet another 
description~\refs{\marcs,\marc} which 
is particularly well adapted to the problem of tachyon 
condensation  \refs{\absapp,\hkm,\bsft}. Consider $2^{m}$ branes with the corresponding gauge 
group $U(2^{m})$ generated by $2^{m}\times 2^{m}$ matrices. These matrices 
can be expanded in terms of $SO(2m)$ gamma matrices,  
\eqn\an{A^{ab}_{\mu} =  \sum_{k=0}^{2m} {1 \over  2k!} A_{\mu}^{I_1 \cdots I_k}
\gamma_{ab}^{I_1 \cdots I_k}~,}
where $\gamma^{I_1 \cdots I_k}$ denote anti-symmetrized products of gamma
matrices with unit weight ({\it e.g.} $\gamma^{12} = \gamma^1 \gamma^2$).  
All that is needed for the expansion \an\ is a representation of the Clifford 
algebra.  So instead of gamma matrices, introduce $2m$ boundary fermion
superfields ${\bf \Gamma}^I = \eta^I + \theta F^I$ with action 
$S = - \int \! d\tau d\theta {1 \over 4}{\bf \Gamma}^I D{\bf \Gamma}^I$.
Canonically quantizing, one arrives at the anti-commutation relations
$\{\eta^I,\eta^J\} = 2\delta^{IJ}$, so $\eta^I$ represent the Clifford
algebra as desired. 
%
%
% so instead of gamma-matrices the generators of 
%$U(2^{m})$ can be realized in terms of boundary fermions 
%$\eta^{I}$, $I=1,\ldots,2m$, by simply replacing $\gamma^{I_1\cdots I_k}$
%with $\eta^{I_1}\cdots \eta^{I_k}$. In the supersymmetric 
%version we introduce $2m$ boundary superfields 
%${\bf \Gamma}^I = \eta^I + \theta F^I$ and
So we write the boundary action for the  non-abelian gauge field as
\eqn\ao{S_A= -\int \! {d\tau} d\theta \, 
\left[ {1 \over 4}{\bf \Gamma}^I D {\bf \Gamma}^I 
 + i \sum_{k=0}^{2m} {1 \over  2k!} A_\mu^{I_1 \cdots I_k}D {\bf X}^\mu
{\bf \Gamma}^{I_1} \cdots {\bf \Gamma}^{I_k} \right]~.}
In this action the correct ordering is enforced by the boundary 
fermions.  Indeed, integrating out ${\bf \Gamma^I}$ using 
the standard formula for a transition amplitude, 
\eqn\uf{\int \! {\cal D}\Phi\,
e^{-S} = {\rm Tr}~ {\rm P} e^{-\int \! d\tau H(\tau)},}
one recovers a path ordered expression. 

The action \ao\ is manifestly invariant under the $U(1)\otimes SO(2m)$ 
gauge transformations
\eqn\ap{\eqalign{ \delta A_\mu & = \partial_\mu \alpha ~,\cr
\delta A_\mu^{IJ} & = \partial_\mu \alpha^{IJ} + 
i(\alpha^{IK}A_\mu^{KJ}- A_\mu^{IK}\alpha^{KJ})~,}}
and with ${\bf \Gamma}^I$, $A_\mu^I$, $A_\mu^{IJK}$, $\ldots$, transforming
in anti-symmetric tensor representations of $SO(2m)$.  The remainder of the
$U(2^m)$ gauge symmetry is realized in a more involved fashion, 
mixing fields with different number of indices. 
%in analogy with the fermionic formulation of the $E_8 \otimes E_8$ heterotic
%string where only an $SO(16) \otimes SO(16)$ subgroup 
%is realized as a symmetry of the classical action.

Actually there is a problem with \ao\ as it stands: 
terms in the action with $k$ odd
 are fermionic rather than bosonic, leading to an incorrect 
algebra. This must be remedied by the introduction of anti-commuting 
cocycle factors. Happily, we will see that this complication is 
absent in the $D\overline{D}$ system. We therefore disregard cocycles 
in the following. Then the boundary fermion representation has an 
important effect on the rules for matrix multiplication. 
Consider a general $2^m \times 2^m$ matrix,
\eqn\wa{M = \pmatrix{A & B \cr C & D} = 
\sum_{k=0}^{2m} M^{I_1 \cdots I_k}\gamma^{I_1 \cdots I_k}.}
We allow each submatrix $A$ $\cdots$ $D$ to be either bosonic or fermionic.
We keep track of this by defining, e.g., $(-)^a$ to be $+1$ if $A$ is
bosonic or $-1$ if $A$ is fermionic.  To this matrix we associate the 
quantity $M = \sum_{k=0}^{2m} M^{I_1 \cdots I_k}{\bf \Gamma}^{I_1 \cdots I_k}$.
Now consider the procedure of multiplying together two matrices expressed
in terms of fermions. We need to enforce the rule that all $\Gamma$'s are
taken to the right of $M^{I_1 \cdots I_k}$'s.   We then  have
\eqn\wb{\eqalign{M M' &= \sum_{k, k'=0}^{2m} 
M^{I_1 \cdots I_k}{\bf \Gamma}^{I_1 \cdots I_k}
M'^{J_1 \cdots J_{k'}}{\bf \Gamma}^{J_1 \cdots J_{k'}} \cr
& = \sum_{k, k'=0}^{2m} (-)^{km'}M^{I_1 \cdots I_k}M'^{J_1 \cdots J_{k'}}
{\bf \Gamma}^{I_1 \cdots I_k} {\bf \Gamma}^{J_1 \cdots J_{k'}}.}}
Here $m'$ is $0$ or $1$ depending on whether 
$M'^{J_1 \cdots J_{k'}}$ is bosonic or fermionic.   So we pick up an 
extra minus sign whenever $M$ is associated with an odd number  
of gamma's and $M'$ is fermionic.  Choosing an off-diagonal basis
for our gamma matrices and
 translating back to 
matrix notation, \wb\ tells us that the correct rule for matrix multiplication
is
\eqn\wc{ MM' =  \pmatrix{A & B \cr C & D} \pmatrix{A' & B' \cr C' & D'}
= \pmatrix{AA' + (-)^{c'} BC' ~~~& AB' +(-)^{d'} BD' \cr && \cr
 DC'+(-)^{a'}CA' ~~~&DD' + (-)^{b'}CB' }.}
This rule is standard in the theory of superconnections \refs{\quillen,\book}.  
It will be
crucial for us later in obtaining the correct tachyon covariant 
derivative.

\subsec{{\bf BSFT for the ${D\overline{D}}$ system}}
There are two unstable D-brane systems in type II string theory:
the ``wrong p'' non-BPS D-branes and the ${D\overline{D}}$ system. 
It is sufficient to consider the string field theory of the 
${D\overline{D}}$ system, 
since the theory of the non-BPS D-brane can then be 
obtained by restricting to couplings which are invariant under 
$(-)^{F_L}$ \FL. Hence we focus on ${D\overline{D}}$.

Consider for definiteness $N$ $D9-\overline{D9}$ pairs in IIB.  This system
has a $U(N) \otimes U(N)$ gauge group with a tachyon transforming in the
$(N,\overline{N})$ representation. The gauge fields and tachyons are 
naturally packaged as $2N \times 2N$ matrices indicating from which open 
string sector the fields arise,
\eqn\aw{\pmatrix{A_\mu^+ & 0 \cr 0& A_\mu^- \cr}, \quad
\quad\quad
 \pmatrix{ 0 & \overline{T} \cr T & 0 \cr}~.}
To write their boundary couplings it is convenient to combine them with
the boundary superfield $D{\bf X}^\mu$ and write
\eqn\ax{ M({\bf X}) = \pmatrix{iA_\mu^+({\bf X}) 
D{\bf X}^\mu & \sqrt{\alpha^\prime}~\overline{T}({\bf X}) \cr  &\cr
\sqrt{\alpha^\prime}T({\bf X}) & iA_\mu^-({\bf X}) D{\bf X}^\mu \cr}.}

Now take $N=2^{m-1}$ so $M$ is a $2^m\times 2^m$ matrix, naturally 
expanded in $SO(2m)$ gamma matrices as in \an,
\eqn\ay{M^{ab} = \sum_{k=0}^{2m} {1 \over  2 k!} M^{I_1 \cdots I_k}
\gamma_{ab}^{I_1 \cdots I_k}~.}
Introducing $2m$ boundary fermion superfields ${\bf \Gamma}^I$ as before, the
boundary interaction is
\eqn\az{S_{{\rm bndy}}= -\int \! {d\tau} d\theta \, 
\left[ {1 \over 4}{\bf \Gamma}^I D {\bf \Gamma}^I 
 +  \sum_{k=0}^{2m} {1 \over  2k!} M^{I_1 \cdots I_k}
{\bf \Gamma}^{I_1} \cdots {\bf \Gamma}^{I_k} \right]~.}
The theory has $U(2^{m-1}) \otimes U(2^{m-1})$ gauge invariance, 
of which a $U(1) \otimes SO(2m)$ subgroup is manifest.   

The lowest component of the superfield $M$ is fermionic and, as for gauge 
fields, the components $M^{I_1 \cdots I_k}$ with odd $k$ have opposite 
statistics, 
{\it i.e.} the lowest component is bosonic. In the present context the 
 bosonic components 
are tachyons so the formalism automatically assigns them the correct 
statistics. Thus we 
do not need to introduce any cocycle factors in the $D{\overline D}$-system. 
In fact, 
it may be useful to realize a non-abelian gauge field on a BPS D-brane
 in terms of a $D{\overline D}$-system 
with $T={\bar T}=A^{-}=0$. This construction has two extra fermions compared 
to the discussion of the pure gauge field around \ao , 
and these additional degrees of 
freedom provide the cocycle factors which were rather 
awkwardly needed in the pure  gauge system.

The upper component of the superfield $\Gamma^{I}=\eta^{I}+\theta F^{I}$ 
is an auxiliary field, as it has no kinetic term. It can be eliminated as 
follows. We write the matrix $M$ as $M=M_{0}+\theta M_{1}$ and carry 
out  the $\theta$-integral in the action \az\ 
\eqn\aza{\eqalign{
S_{{\rm bndy}} &= -\int \!\!d\tau \left[ {1\over 4}\dot{\eta}^{I}\eta^I
+ {1\over 4}F^{I}F^{I} +\right. \cr
& \quad\quad \left. \sum_{k=0}^{2m} {1\over 2k!}
\left(M_{1}^{I_1 \cdots I_k}\eta^{I_1} \cdots \eta^{I_k} 
-(-1)^k kM_{0}^{I_1 \cdots I_k}F^{I_{1}}\eta^{I_2} \cdots 
\eta^{I_k} \right)\right]~.}
}
The Gaussian integral over the $F^{I}$ can now be carried out. The result 
is a term $-{1\over 4}F^{I}F^{I}$ under the integral, where
\eqn\fondef{
F^{I} = \sum_{k=1}^{2m} {(-1)^k\over (k-1)!}
M_{0}^{II_{2} \cdots I_k}\eta^{I_2} \cdots \eta^{I_k}~. 
}
To write the result in a more compact form, consider a general 
$2^m\times 2^m$ 
matrix, represented in terms of $SO(2m)$ matrices as in 
\an . Simplifying the products of gamma matrices using the
Clifford algebra
yields an identity of the form
\eqn\mthe{\eqalign{ 
&~~~\sum_{k,k'=0}^{2m}{1\over 2k! 2k'!}
\Phi^{I_{1} \cdots I_k} \gamma^{I_1} \cdots  \gamma^{I_k}\Phi^{J_{1} 
\cdots J_{k'}} \gamma^{J_1} \cdots  \gamma^{J_{k'}} \cr 
&=  
\sum_{k,k'=0}^{2m}{(-)^{k+k'}\over 4(k-1)!(k'-1)!}\Phi^{I I_{2} 
\cdots I_k} \gamma^{I_2} \cdots  \gamma^{I_k}\Phi^{I J_{2} \cdots J_{k'}} 
\gamma^{J_2} \cdots  \gamma^{J_{k'}}
~+ ~{\rm higher~contractions}
~.}}
%\eqn\mthe{\eqalign{
%\Phi^{2} &= 
%\left(\sum_{k=0}^{2m}{1\over 2k!}\Phi^{I_{1} \cdots I_k} \gamma^{I_1} \cdots 
%\gamma^{I_k}\right)^2 \cr 
%&= 
%\left(
%\sum_{k=1}^{2m}{(-1)^k\over 2(k-1)!}\Phi^{II_{2} \cdots I_k}\gamma^{I_2} 
%\cdots 
%\gamma^{I_k}\right)^{2}+ {\rm higher~contractions}
%~.}}
Using this identity for $\Phi=M_0$ and temporarily ignoring 
 the higher contractions we find the action
\eqn\azb{
S_{{\rm bndy}} = -\int\! d\tau \left[ {1\over 4}\dot{\eta}^{I}\eta^I
+\sum_{k=0}^{2m} {1\over 2k!}
(M_{1}-M_{0}^{2})^{I_1 \cdots I_k}\eta^{I_1} \cdots \eta^{I_k} \right]~.
}
Here $M_0^2$ is defined using the matrix multiplication rule of \wc\
with $a^\prime =d^\prime =1$ and $b^\prime =c^\prime =0$ 
(becaue off-diagonal entries of $M_0$ are bosonic while diagonal entries
are fermionic). 

In the Gaussian computation yielding $-{1\over 4}F^{I}F^{I}$ the interacting 
fermions were treated na\"{\i}vely, omitting the presence of divergences
when two fermions coincide. In more precise computations 
there are additional contractions between the fermions in the
two $F^I$'s. 
In \marc\ it is shown 
that these contractions exactly match the higher contractions in 
\mthe, rendering \azb\ the complete result. The $M_0^2$ terms in 
\azb\  have a nice interpretation: they give the commutator terms 
in the non-abelian field strengths of $A^+$ and $A^-$, as well as 
in the gauge covariant derivatives of the tachyon field. 
These results were guaranteed by gauge invariance; they are 
nevertheless nontrivial to recover because not all of the gauge invariance is 
manifest in the  present formalism.
\newsec{Explicit Computations}
In this section we make these general considerations explicit for 
$m=1$, corresponding to a single $D9-\overline{D9}$ pair.
In this case gamma matrices reduce to Pauli matrices, and the 
expansion \ay\ takes the form
\eqn\ba{M = {i\over 2} A_\mu D{\bf X}^\mu {\bf 1} +
{1 \over 2}\sqrt{\alpha^\prime} T^I \sigma_I + 
{i \over 4}A_\mu^{IJ}D{\bf X}^\mu\sigma_{IJ}, \quad~
I,J =1,2~,}
with $\sigma_{IJ} = [\sigma_I,\sigma_J]/2$ and 
\eqn\bb{\eqalign{ A_\mu^{\pm} &= {1 \over 2}(A_\mu \pm iA_\mu^{12})~,\cr
T &= {1 \over 2}(T^1 + i T^2)~.}}
Therefore the boundary interaction is
\eqn\bc{S_{\rm bndy} = -\int \! {d\tau} d\theta \, \left[
{1\over 4} {\bf \Gamma}^I D {\bf \Gamma}^I + 
{i\over 2} A_\mu D{\bf X}^\mu  + 
 {1 \over 2}\sqrt{\alpha^\prime}T^I {\bf \Gamma}^I 
+ {i \over 4} A_\mu^{IJ} D{\bf X}^\mu {\bf \Gamma}^I {\bf \Gamma}^J
\right]~.}
This action  exhibits the full $U(1)\otimes U(1)$ gauge 
symmetry. Expanding \bc\ in components and integrating out the 
auxilliary fields $F^I$ (recall ${\bf \Gamma}^I = \eta^I + 
\theta  F^I$) yields
\eqn\bd{\eqalign{S_{{\rm bndy}} = -\int \! {d\tau} \left[  
 -{\alpha^\prime\over 4}T^I T^I  \right. &
+  {1 \over 4} \dot{\eta}^I \eta^I +
 {\alpha^\prime\over 2}D_\mu T^I \psi^\mu \eta^I 
+{i\over 2}(\dot{X}^\mu A_\mu +
{1 \over 2}\alpha' F_{\mu\nu}\psi^\mu \psi^\nu)   \cr
&\left.
\!\!\!\!\!\!\!\!\!\!\!\!\!\!\!\!\!\!\!\!\!\!
+{i\over 4}(\dot{X}^\mu A^{IJ}_\mu 
+{1 \over 2}\alpha'F^{IJ}_{\mu\nu}\psi^\mu \psi^\nu)
\eta^I\eta^J  \right]~,}}
where the derivative of the tachyon comes out correctly covariantized,
\eqn\be{D_\mu T^I = \partial_\mu T^I - i A^{IJ}_\mu T^J~.}
\bd\ is 
in agreement with the general results in the previous section.

The interactions in \bd\ are nontrivial so the corresponding partition 
function cannot be computed explicitly in its entirety. In the following 
we consider various special cases.
 
\subsec{{\bf Tachyon Condensation on $D\overline{D}$.}}
To begin, we set the gauge fields to zero, $A^+_\mu = A^-_\mu=0$.
Then it is simple to integrate out $\eta^I$, with the result
\eqn\bff{S_{{\rm bndy}} = {\alpha^\prime\over 4} \int \! {d\tau} \left[ 
T^I T^I +{\alpha'}(\psi^\mu \partial_\mu T^I){1 \over \partial_\tau }
(\psi^\nu \partial_\nu T^I) \right]~.}
The operator $1/\partial_\tau$ is defined by
\eqn\bg{{1 \over \partial_\tau }f(\tau) = {1 \over 2} \int \! d\tau' \,
\epsilon(\tau -\tau')f(\tau')~,}
where $\epsilon(\tau)$ is $+1$ or $-1$ for positive or negative $\tau$. 

For constant tachyon we have simply
\eqn\cnsta{
\Psi_{\rm bndy} = e^{-S_{\rm bndy}} = e^{-2\pi\alpha^\prime T{\bar T}}~.
}
The disc partition function is obtained by projecting this onto the
bulk wave functional corresponding to \ag , and integrating over all
fields. No further tachyon dependence is introduced in this process
so we learn that the tachyon potential for the $D\overline{D}$ 
system is 
\eqn\bh{V(T,\overline{T}) = 
2T_{D9}\, e^{-2\pi\alpha^\prime T \overline{T}}~,}
where we fixed the overall normalization by hand, though this  presumably can
be verified independently  as in \sennor .  

Next we turn to spatially dependent tachyon configurations. Linear 
tachyon profiles are singled out as leading to free worldsheet theories.  
By a combination of spacetime and gauge rotations we can bring $T^I$ 
to the form
\eqn\bi{\sqrt{\alpha^\prime}T^I = u^I X^I~.}
Substituting  the mode expansions \d\ and \af\ into \bff\ and combining with
\ag\ gives the action
\eqn\bj{S_{{\rm bulk}}+S_{{\rm bndy}} = 
2\pi \alpha^\prime T{\bar T} + \sum_{I=1}^2 \left[
{1 \over 2} \sum_{n=1}^\infty(n+\pi{\alpha'} {y^I})X_{-n}^I X_n^I
+ i\sum_{r={1 \over 2}}^\infty(1+\pi{\alpha'} 
{y^I\over r})\psi^I_{-r}\psi^I_r \right]~.} 
The first term is simply the zero-mode part of \bi\ 
and we defined \eqn\bk{y^I = (u^I)^2~.}
The partition function is
\eqn\bl{\eqalign{Z(y^I) &= \int\! {d^{10}X_0 \over (2 \pi \alpha')^5}
\prod_{I=1}^2 \left(\prod_{n=1}^\infty 
{dX^I_n dX^I_{-n} \over 4\pi}  \prod_{r={1 \over 2}}^\infty 
d \psi^I_r d\psi^I_{-r}\right)
e^{-(S_{{\rm bulk}} + S_{{\rm bndy}})}\cr & \cr
&= \int \!{d^{10}X_0 \over (2 \pi \alpha')^5}\,
e^{-2\pi\alpha^\prime T{\bar T}}\prod_{I=1}^2 
{\prod_{r={1 \over 2}}^\infty \left(1+\pi{\alpha'} {y^I\over r}\right) \over
\prod_{n=1}^\infty (n+\pi{\alpha'} {y^I})}~.
}}
The bosonic product in the denominator was computed in \o\ and
the fermionic product is similarly
\eqn\bm{\prod_{r={1 \over 2}}^\infty \left(r+\pi{\alpha'} {y^I}\right)
= \prod_{n=1}^\infty  ( n +\pi{\alpha'} {y^I} - {1 \over 2})
= {\sqrt{2\pi} \over \Gamma\left(\pi{\alpha'} {y^I} +{1\over 2}\right)}
= {4^{\pi{\alpha'} {y^I}} \over \sqrt{2}}{\Gamma({\pi\alpha'} 
{y^I}) \over 
\Gamma(2\pi\alpha' y^I)}~.}
Defining the function
\eqn\fdeg{
F(x) = \sqrt{2\pi}~{\prod_{r={1\over 2}}^\infty (1 + {x\over r})\over
\prod_{n=1}^\infty (n+x)} = {4^x x \Gamma(x)^2\over 2\Gamma(2x)}~,
}
our result for the partition function becomes
\eqn\bn{Z(y^I) = 2T_{D9} \int\! d^{10}X_0 \,
~e^{-2\pi\alpha^\prime T{\bar T}} 
\prod_{I=1}^2 F(\pi\alpha^\prime y^I)~.}
The overall normalization was fixed by comparison with \bh .

The partition function \bn\ gives the spacetime action evaluated on 
linear tachyon profiles. \bi\ shows that it can be written in 
terms of the tachyon by the substitution 
$y^I\to \alpha^\prime (\partial_I T^I)^2 $. This result gives 
an expression for the action to all orders in 
derivatives. However, a significant ambiguity remains: 
any term with at least second derivatives acting on $T$ can be added.
At quadratic order in derivatives the result is unambiguous:
\eqn\bq{S(T,\overline{T}) \simeq  2T_{D9} \int \! d^{10}x \, 
e^{-{2\pi\alpha^\prime}T\overline{T}} \left[ 1 +  8\pi (\alpha')^2 
\ln(2) \partial^\mu \overline{T} \partial_\mu T + \cdots \right]~.}
We used the expansion
\eqn\bp{F(x) \simeq 1+ 2 \ln (2) x + O(x^2),
\quad~ x \rightarrow 0~.} 
%We used the limiting behaviors of $F$ are
%\eqn\bp{F(x) \sim \left\{\matrix{1+ 2 \ln (2) x + O(x^2)~, 
%& \quad x \rightarrow 0~, \cr  & \cr 
%\sqrt{ \pi x } + O(x^{-{1 \over 2}})~, & \quad x \rightarrow \infty~.
% \cr 
%}\right.
%}

Next we turn to the description of lower dimensional D-branes as solitons
on the $D9-\overline{D9}$ system. According to the conjectures of 
Sen (for a review see \senrev ), a kink  represents a non-BPS 
$D8$-brane and a vortex represents a BPS $D7$-brane.  There are three 
fixed points of the RG flow depending on whether zero, one, or both 
of the $y^I$ are taken to infinity, and these represent 
the $D9-\overline{D9}$, the non-BPS $D8$-brane, and the BPS $D7$-brane, 
respectively. A single nonzero $y^I$ gives a tachyon profile $T \sim x^1$, 
which indeed describes a kink; and two nonzero $y^I$'s 
gives $T \sim x^1 + i x^2$ which describes a vortex. To 
compute the tension of these solitons we simply evaluate \bn\ at 
the endpoint of the RG flow using the limiting behavior
\eqn\bpa{F(x) \simeq
\sqrt{ \pi x } + O(x^{-{1 \over 2}})~, \quad x \rightarrow \infty~.
}
We find:
\item{(1)} A non-BPS D8-branes corresponds to
$y^1 = \infty$ and $y^2=0$. This gives
\eqn\bsa{\eqalign{Z(y^1,0) &= 2T_{D9} \int\! d^{10}X_0 \, 
e^{-{\pi \over 2}y^1 (X_0^1)^2}
F({\pi \alpha'} y^1) \cr
&= 2T_{D9} \int\! d^{9}X_0 \, 
\sqrt{{2 \over y^1}}
F({\pi \alpha'} y^1) \cr 
& \longrightarrow~  2\pi \sqrt{2 \alpha'}~ T_{D9} \int\! d^{9}X_0~,}}
which correctly identifies the tension as $T_8 
= \sqrt{2}(2 \pi \sqrt{\alpha'})T_{D9}=\sqrt{2} T_{D8} .$
\item{(2)} The BPS D7-brane corresponds to $y^1 = y^2=\infty$.
This gives
\eqn\bta{\eqalign{Z(y^1,y^2) &= 2T_{D9} \int\! d^{10}X_0 \, 
e^{-{\pi \over 2}[y^1 (X_0^1)^2+y^2 (X_0^2)^2]}
F({\pi\alpha'} y^1)F({ \pi\alpha'} y^2) \cr
&= 2T_{D9}\int\! d^{8}X_0 \, \sqrt{{2 \over y^1}}
F({ \pi\alpha'} y^1)\sqrt{{2 \over y^2}}
F({ \pi \alpha'} y^2)~ \cr
& \longrightarrow~  4 \pi^2\alpha' T_{D9} \int\! d^{9}X_0~,}}
which correctly gives the tension as
$T_{D7} = (2 \pi \sqrt{\alpha'})^2 T_{D9}$.

%\vskip 0.5cm
%\noindent
%$\underline{{\bf y^1 = \infty,~  y^2=0}}$
%\vskip 0.2cm
%\eqn\bs{\eqalign{Z(y^1,0) = 2T_{D9} \int\! d^{10}X_0 \, 
%e^{-{\pi \over 2}y^1 (X_0^1)^2}
%F({\pi \alpha'} y^1) &= 2T_{D9} \int\! d^{9}X_0 \, 
%\sqrt{{2 \over y^1}}
%F({\pi \alpha'} y^1) \cr & \cr
%&\longrightarrow~  2\pi \sqrt{2 \alpha'}~ T_{D9} \int\! d^{9}X_0,}}
% which correctly identifies the tension of the 
%non-BPS $D8$-brane as $T_8 
%= \sqrt{2}(2 \pi \sqrt{\alpha'})T_{D9}=\sqrt{2} T_{D8} .$
%\vskip 0.5cm
%\noindent
%\item{$\underline{{\bf y^1 = \infty,~  y^2=\infty}}$}
%\eqn\bt{\eqalign{Z(y^1,y^2) &= 2T_{D9} \int\! d^{10}X_0 \, 
%e^{-{\pi \over 2}[y^1 (X_0^1)^2+y^2 (X_0^2)^2]}
%F({\pi\alpha'} y^1)F({ \pi\alpha'} y^2) \cr & \cr
%&= 2T_{D9}\int\! d^{8}X_0 \, \sqrt{{2 \over y^1}}
%F({ \pi\alpha'} y^1)\sqrt{{2 \over y^2}}
%F({ \pi \alpha'} y^2)~ 
%\longrightarrow~  4 \pi^2\alpha' T_{D9} \int\! d^{9}X_0,}}
%which correctly gives the $D7$-brane tension as
%$T_{D7} = (2 \pi \sqrt{\alpha'})^2 T_{D9}$.
\noindent
Higher codimension branes can be described similarly. The 
details of this generalized construction is discussed in sec 5.4.

\subsec{{\bf Gauge fields on the  $D\overline{D}$ system}}
We now consider simple examples with both tachyons and gauge fields 
on the $D\overline{D}$ system. To do so, we return to the boundary 
action \bd.  

Setting $A^{IJ}_\mu =0$, $F_{\mu\nu}={\rm constant}$, 
$T^I={\rm constant}$ leads to the action
\eqn\btb{\eqalign{
S &= 2 T_{D9}\int\! d^{10}x\, 
e^{- {\pi\alpha^\prime\over 2}T^I T^I} {\cal L}_{BI}(F/2) \cr
& =  \int\! d^{10}x\, 
V(T,{\bar T}){\cal L}_{BI}( F^+)~,
}}
where $V$ is the tachyon potential \bh . 
When the two gauge fields are identical, $F^+=F^-$,
we thus find a  Born-Infeld action times
 an overall factor equal to the tachyon potential \senuniv .

Next, consider $F_{\mu\nu}$ and  $F^{IJ}_{\mu\nu}$ constant and $T^I =0$.   
First integrate out $\eta^I$ using
\eqn\bt{ \int \! D\eta \, e^{- \int \! {d\tau}\, \left[
{1 \over 4} \dot{\eta}^I\eta^I  +  
{i \over 2}N(\tau) \epsilon^{IJ} \eta^I \eta^J \right]}
= e^{ \int \! {d\tau}\, N(\tau)}  + 
e^{- \int \! {d\tau}\, N(\tau)}~.} 
The partition function then becomes 
\eqn\bu{Z(A^+,A^-) = \int\! {\cal D}X {\cal D}\Psi \, e^{-S_{{\rm bulk}}}\left[
e^{-S^+_{{\rm bndy}}} +  e^{-S^-_{{\rm bndy}}} \right],}
with
\eqn\bv{S^\pm_{{\rm bndy}} = -{i} \int \! {d\tau}\,
\left[ 
\dot{X}^\mu A^\pm _\mu 
+{1 \over 2}{\alpha'}F^\pm_{\mu\nu}\psi^\mu \psi^\nu \right].}
Therefore, the partition function for this background is a sum of two 
Born-Infeld actions,
\eqn\bw{ Z(A^+,A^-) = T_{D9} \int \! d^{10} x\,  \left[
{\cal L}_{BI}(F^+) + {\cal L}_{BI}(F^-)  \right].}  
This is correct, since for vanishing tachyon the gauge fields on the 
two D-branes are  decoupled from one another. 

\subsec{\bf Mixing of Gauge Fields and Tachyons}
Finally, we turn to the more nontrivial case of constant and nonzero
$F_{\mu\nu}$,  $F^{IJ}_{\mu\nu}$, and $T^I$.  In this case we will work
out the partition function perturbatively in $A^{IJ}_{\mu}$, which 
corresponds to expanding in $D_\mu T^I$ and $F^{IJ}_{\mu\nu}$. From 
\bd\ it follows that each such term has the tachyon dependence  
$e^{-{\pi\alpha^\prime \over 2}T^I T^I}$ 
times a polynomial in $T^I T^I$; in particular,
all the terms vanish in the closed string vacuum $T^I T^I \rightarrow \infty$.
Now let's work out the explicit terms quadratic in field strengths. 
Integrating out $\eta^I$ at this order yields the partition function
\eqn\bx{Z(T,A^+,A^-) = \int\! DX D\psi \, e^{-S_{{\rm bulk}}}\left[
e^{-S^+_{{\rm bndy}}} +  e^{-S^-_{{\rm bndy}}} \right],}
with 
\eqn\by{S^\pm_{{\rm bndy}} =  \int \! d\tau\,
\left[ {\alpha^\prime \over 4}T^I T^I + 
{(\alpha')^2 \over 4}(D_\mu T^I \psi^\mu) 
{1 \over \partial_\tau}(D_\nu T^I \psi^\nu)  
- i(\dot{X}^\mu A^\pm _\mu +
{\alpha' \over 2}F^\pm_{\mu\nu}\psi^\mu \psi^\nu) \right].}
We stress that \by\ is only correct to order $A^2$.   In terms of 
$A^\pm$ we have 
\eqn\bz{D_\mu T^I = \partial_\mu T^I - (A^+_\mu - A^-_\mu)
\epsilon^{IJ}T^J~.}

We write the background as
\eqn\ca{A_\mu^\pm = - {1 \over 2}F^{\pm}_{\mu\nu}X^\nu, \quad \quad
F^{\pm}_{\mu\nu},~ T^I = ~{\rm constant}~,}
hence
\eqn\cb{D_\mu T^I = {1\over 2}\epsilon^{IJ}T^J(F^{+}_{\mu\nu}-F^{-}_{\mu\nu})
 X^\nu~.
}
At order $A^2$ we can separate \bx\ into two terms, $Z= Z^{(0)} + Z^{(1)}$, 
corresponding to 
expanding \bx\ to zeroth and first order in $(D_\mu T^I \psi^\mu) 
{1 \over \partial_\tau}(D_\nu T^I \psi^\nu)$.  The zeroth order term 
is \bw\ and
%\eqn\cc{Z^{(0)} =   T_{D9} \int \! d^{10}x \, 
%e^{-{\pi\alpha^\prime\over 2}T^I T^I}  \left[
%{\cal L}_{BI}(F^+) + {\cal L}_{BI}(F^-)  \right]~.
%} 
the first order term is 
\eqn\cd{Z^{(1)} =  {(\alpha^\prime)^2\over 16} 
e^{-{\pi\alpha^\prime\over 2}T^I T^I}T^J T^J
(F^+_{\mu\alpha}-F^-_{\mu\alpha})(F^+_{\nu\beta}-F^-_{\nu\beta})
\!\int \! d\tau d\tau'\, \epsilon(\tau-\tau')  
\langle X^\alpha(\tau)X^\beta(\tau')\psi^\mu(\tau)\psi^\nu(\tau')\rangle~,}
where we used \bg, and where
\eqn\ce{
\langle X^\alpha (\tau)X^\beta (\tau^\prime)
\psi^\mu (\tau)\psi^\nu (\tau^\prime)\rangle =
\int \! {\cal D}X {\cal D}\psi\, e^{-S_{{\rm bulk}}}
X^\alpha (\tau)X^\beta (0)\psi^\mu (\tau)\psi^\nu (0)~.}
Separating out the $X$ zero mode, we write
\eqn\cf{\eqalign{ \langle X^\alpha (\tau)X^\beta (\tau')\rangle &= 
\delta^{\alpha\beta}
G(\tau,\tau') + X_0^\alpha X_0^\beta~, \cr
\langle \psi^\mu(\tau)\psi^\nu(\tau')\rangle &= 
\delta^{\mu\nu}K(\tau,\tau')~,}}
where the correlators
\eqn\g{\eqalign{G^{-1}(\tau,\tau^\prime) &= 
{1\over 4\pi^2 \alpha'}\sum_{n=1}^{\infty} 
n \cos n(\tau-\tau^\prime)~,\cr
K^{-1}(\tau,\tau^\prime) &= - 
{1 \over 4\pi^2}\sum_{r={1 \over 2}}^{\infty}
\sin r (\tau-\tau^\prime)~,}
}
are defined so that the bulk action \ag\ reads
\eqn\ad{ S_{\rm bulk} =  
\int\! d\tau d\tau^\prime \, \left[
X^\mu(\tau) G^{-1}(\tau,\tau^\prime) X_\mu(\tau^\prime) +  
\psi^\mu(\tau) K^{-1}(\tau,\tau^\prime)\psi_\mu(\tau^\prime)\right]~.
}
Now insert \cf\ into \cd.  The contribution with an explicit dependence on the 
zero modes $X_0$ combine with the earlier result \bq\ to provide the
gauge covariant tachyon kinetic term $(D_\mu T^I)^2$.  The remainder
contributes to the gauge kinetic terms as
\eqn\cg{ {\beta \over 4}(\alpha^\prime)^2
e^{-2{\pi\alpha^\prime}T {\bar T}} T{\bar T}
(F^+_{\mu\nu}-F^-_{\mu\nu})^2~,
}
where we defined
\eqn\ch{\beta = \int \! d\tau d\tau^\prime\, \epsilon(\tau-\tau') 
G(\tau,\tau')K(\tau,\tau^\prime)~.}
Combining this result with the order $F^2$ expansion of \bw\ gives our 
result for the partition function at this order
\eqn\ci{\eqalign{Z(T,\overline{T},A^+,A^-) & = 2T_{D9} \int \! d^{10}x \, 
e^{-2\pi\alpha^\prime T\overline{T}} \left[ 1 + 8\pi \alpha' \ln(2) 
D^\mu \overline{T} D_\mu T \right.  \cr
&
%\!\!\!\!\!\!\!\!\!\!\!\!\!\!\!\!\!\!\!\!\!\!\!\!\!\!\!\!\!\!\!\!\!\!\!
%\!\!\!\!\!\!\!\!\!\!\!\!\!\!\!\!\!\!\!\!\!\!\!\!\!\!\!\!\!\!\!\!\!\!\!
%\!\!\!\!\!\!\!\!\!\!\!\!\!\!\!\!\!\!\!\!\!\!\!\!\!\!\!\!\!\!\!\!\!\!\!
+\left. {(2\pi\alpha')^2 \over 8}(F^+_{\mu\nu})^2 
+{(2\pi\alpha')^2 \over 8}(F^-_{\mu\nu})^2 
+ {\beta \over 8}(\alpha^\prime)^2
T\overline{T} (F^+_{\mu\nu}-F^-_{\mu\nu})^2
  \right]~.}}
The actual numerical value of $\beta$ could be computed from 
\ch , after regularization and renormalization.  

\subsec{\bf Non-BPS Branes}
Before ending this section let us comment on non-BPS D-branes.
They are defined as projections by $(-1)^{F_L}$ of 
the $D{\bar D}$ system \FL, so are  included as special
cases of our formalism. Results for a single non-BPS brane 
can be obtained from the boundary interaction \bd\ after the 
following substitutions:
\item{(1)} 
The tachyon is taken to be real. So  take $T_2=0$ and
thus $T={\bar T}={1\over 2}T_1$.
\item{(2)}
The two gauge fields are identified, 
$F^+ = F^- = F_{{\rm non-BPS}~D}$.
\item{(3)} 
After the above restrictions, the fermion $\eta^{2}$ decouples.  One should
not perform the path integral over this fermion. This corresponds to
changing the overall normalization from $2T_{D9}$ to $\sqrt{2}T_{D9}$,
which is the correct tension of a non-BPS D9-brane.
\bigskip
\noindent
For example, the tachyon action \bn\ translates to the non-BPS action
\eqn\br{S(T) = \sqrt{2}T_{D9} \int \!d^{10}x \,
e^{-2\pi\alpha^\prime T^2} 
F(2{\pi (\alpha')^2} \partial^\mu T \partial_\mu T )~.}
This is the same result found in \bsft , after the 
identification $T_{\rm there}^2 = 8\pi\alpha^\prime T_{\rm here}^2$.
In sec 5.5 we use the same identifications to compute the coupling 
to RR fields.

\newsec{Couplings to RR-fields}
In this section we derive the Chern-Simons 
couplings between unstable brane systems 
and background RR-fields. The computations reduce to integrals over
 fermion zero-modes. The results are presented in (5.34) and (5.48).

\subsec{\bf Wave Functionals Revisited}

So far we have studied the action for a D-brane in the 
closed string NS-NS vacuum.   To include closed string excitations
we should compute the path integral on the disk with insertions of
bulk vertex operators.   By first performing the path integral over
bulk fields we obtain a bulk wave functional representing the closed 
string background. Projecting  against the 
boundary wave functional then yields their coupling to open strings.
In this way we arrive at a prescription for coupling off-shell open
strings to on-shell closed strings. 

The interest in this section is to study couplings to RR-fields. As the
starting point we need the bulk wave functional of the RR-vacuum. We begin
by developing the wave functional formalism a little further before
taking couplings to RR-fields into account.

The bulk wave functionals are the state vectors of the closed string, and
operators acting on them 
 form representations of the closed string operator algebra. An
explicit construction of the bosonic operators
in terms of  the modes $X_n^\mu$ is
\eqn\ta{
\alpha^\mu_{n} = 
-{in\over 2}X^\mu_{-n} - i {\partial\over\partial X^\mu_{n}}~,~~~~~
{\tilde\alpha}^\mu_{n} = -{in\over 2}X^\mu_{n} - i 
{\partial\over\partial X^\mu_{-n}}~.
}
These operators indeed satisfy the standard commutation relations
\eqn\tb{
[ \alpha_{n}^\mu, \alpha_{m}^\nu ] = n\delta^{\mu\nu}\delta_{n+m}~,~~~~~
[ \tilde{\alpha}_{n}^\mu, \tilde{\alpha}_{m}^\nu ] = 
n\delta^{\mu\nu}\delta_{n+m}~,~~~~~
[ \alpha_{n}^\mu, \tilde{\alpha}_{m}^\nu ] = 0~.
}
The zero-mode operators are identical, 
$\alpha_0^\mu=\tilde{\alpha}_0^\mu$, as they should be.

These considerations are independent of the specific wave 
functional being considered. In vacuum, the bosonic 
part of the wave functional is \bulksv\
\eqn\tc{
\Psi_{\rm bulk}^{\rm bos}= {\cal N}_{\rm bos}\exp \left[
-{1\over 2}\sum_{n=1}^{\infty}nX_{-n}^\mu X_{n}^\mu\right]~.
}
In this particular state 
\eqn\vacrel{
\alpha_{n}^\mu\Psi_{\rm bulk}^{\rm bos}=
\tilde{\alpha}_{n}^\mu\Psi_{\rm bulk}^{\rm bos} = 0~,
}
for $n\geq 0$. This confirms that the state is the bosonic vacuum. 

The next step is to include fermions. In the NS sector
the  fermion field 
$\psi$ has modes $\psi_{r}$, $r\in\ZZ+{1\over 2}$. Functionals
of such fields are acted on by the closed string
fermion operators
\eqn\td{
\beta_{r}^\mu = {1\over\sqrt{2i}}(\psi_{-r}^\mu +i 
{\partial\over\partial\psi_{r}^\mu})~,~~~~~
{\tilde\beta}_{r}^\mu = {\sqrt{i\over 2}}(\psi_{r} -
i{\partial\over\partial\psi^\mu_{-r}})~,
}
satisfying the standard anti-commutation relations 
\eqn\te{
\{ \beta_{r}^\mu, \beta^\nu_{s} \} = \delta^{\mu\nu}\delta_{r+s}~,~~~~~
\{ \tilde{\beta}_{r}^\mu, \tilde{\beta}_{s}^\nu \} = 
\delta^{\mu\nu}\delta_{r+s}~,~~~~~
\{ \beta_{r}^\mu, \tilde{\beta}_{s}^\nu \} = 0~.
}
Again, consider as a definite example the vacuum wave functional 
\eqn\tf{
\Psi_{\rm bulk}^{NS-NS} = {\cal N}_{NS-NS}\exp\left[ 
-{1\over 2}\sum_{n=1}^{\infty}nX_{-n}^\mu X_{n}^\mu
 -i\sum_{r={1\over 2}}^{\infty}\psi_{-r}^\mu\psi_{r}^\mu\right]~.
}
In this state
\eqn\fvacrel{
\beta_{r}^\mu\Psi_{\rm bulk}^{NS-NS}=
\tilde{\beta}^\mu_{r}\Psi_{\rm bulk}^{NS-NS} = 0~,
}
for $r>0$ and \vacrel\ remains satisfied. This confirms that the 
state is the NS-NS vacuum. Our convention for fermionic derivatives is 
that they act from the left. 

We are now ready to determine the RR vacuum. The  fermions now have 
integer modes $\psi_{n}^\mu$, 
$n\in\ZZ$ and realize the closed string algebra 
\eqn\tf{
\{ \beta_{n}^\mu, \beta_{m}^\nu \} = \delta^{\mu\nu}\delta_{n+m}~,~~~~~
\{ \tilde{\beta}_{n}^\mu, \tilde{\beta}_{m}^\nu \} = 
\delta^{\mu\nu}\delta_{n+m}~,~~~~~
\{ \beta_{n}^\mu, \tilde{\beta}_{m}^\nu \} = 0~,
}
through
\eqn\tg{
\beta_{n}^\mu = {1\over\sqrt{2i}}(\psi_{-n}^\mu +i 
{\partial\over\partial\psi_{n}^\mu})~,~~~~~
{\tilde\beta}_{n}^\mu = \sqrt{i\over 2}(\psi_{n}^\mu -i 
{\partial\over\partial\psi_{-n}^\mu})~.
}
A fundamental aspect of the RR-sector is the role played by the algebra 
of fermion zero-modes. Spinorial representations 
of the Lorentz group are realized by the identification 
$\Gamma^{\mu}=\sqrt{2}\beta_{0}^{\mu}$ and similarly for the right 
movers, $\tilde{\Gamma}^{\mu}=\sqrt{2}\tilde{\beta}_{0}^{\mu}$. 
These operators act on wave functionals of the zero-modes $\psi_{0}^{\mu}$. 
 In the RR-vacuum the wave functional is thus
\eqn\th{
\Psi_{\rm bulk}^{RR} = {\cal N}_{RR}\exp\left[ 
-{1\over 2}\sum_{n=1}^{\infty}nX_{-n}^\mu X_{n}^\mu
-i\sum_{n=1}^{\infty}\psi^\mu_{-n}\psi^\mu_{n}\right]
\sum_{p~ {\rm odd}}
{(-i)^{9-p\over 2}\over (p+1)!}
C_{\mu_{0}\cdots\mu_{p}}\psi^{\mu_{0}}_{0}
\cdots \psi^{\mu_{p}}_{0}~,
}
where the $C_{\mu_{0}\cdots\mu_{p}}$ are numerical coefficients that 
will be identified with RR-potentials momentarily. 
We chose for definiteness the type IIB GSO projection, 
acting on the zero-mode vacua by 
an even number of $\Gamma^\mu$ as well as an even number of 
$\tilde{\Gamma}^\mu$. In type IIA 
%$\Gamma^{\mu}$ or ${\tilde\Gamma}^\mu$ is applied;
 the result is the same 
except that  $p$ is restricted to even integers.

It is instructive to compare these considerations with standard 
worldsheet technology.
By the state-operator correspondence we can associate each wave functional
on the boundary of the disk to a vertex operator inserted at the 
center of the disk. The RR-vacuum wave functional \th\ corresponds to an insertion 
of
\eqn\ti{
{\cal\nu}_{RR}^{(-{1\over 2},-{3\over 2})} = 
S^{a}C_{ab}{\tilde S}^{b}~
e^{-{1\over 2}\phi(0)}e^{-{3\over 2}\tilde{\phi}(0)}~.
}
 The RR-potential 
is written here in a spinorial form with the component 
expansion\foot{This 
is for type IIB. For type IIA the corresponding spinors have 
opposite chirality; the potential is $C_{a\dot{b}}$.}
\eqn\rrc{
C_{ab} = \sum_{p~{\rm odd}}
{1\over (p+1)!}C_{\mu_{0}\cdots\mu_{p}}(\Gamma^{\mu_{0}\cdots\mu_{p}})_{ab}~.
}
After the vertex insertion the 
fermion field $\psi$ becomes integer moded. Additionally, there is an 
overall factor related to the RR-potential. The conclusion is that the 
coefficients $C_{\mu_{0}\cdots\mu_{p}}$ in \th\ can be identified with the 
RR-potential. In the construction of \th\ normalizations were determined 
using the correspondence 
$\Gamma^{\mu}\sim\sqrt{2}\beta_{0}^{\mu}\sim 
i^{1\over 2}\psi_{0}^{\mu}$ 
when acting on the zero-mode vacuum (and similarly for right movers). 
The coefficients $C_{\mu_{0}\cdots\mu_{p}}$ 
in \th\ are therefore correctly normalized. In 
other words, the overall factor ${\cal N}_{RR}$ is independent of $p$. 
It is possible to determine the numerical value of ${\cal N}_{RR}$ explicitly 
by computations familiar from the boundary state formalism (see {\it e.g.} 
\bsrefs ). In the following it will be fixed by requiring the 
correct D9-brane charge.  

The operator \ti\ is written in the $(-{1\over 2},-{3\over 2})$ picture. 
The total number of superconformal ghosts is thus $-2$, saturating the 
superconformal Killing symmetries on the disc. As usual, on the disc the 
ordinary ghosts are taken care of by fixing the bulk vertex operator 
at the origin. This leaves one CKV, corresponding to the azimuthal symmetry 
of the disc boundary. In principle this means one boundary operator must 
be fixed, but integrating instead over all boundary operators, as we 
will find convenient, simply 
results in an overall numerical factor that can be 
ignored. The partition sums computed here are thus interpretable as generating 
functionals of  string amplitudes. In the previous sections ghosts and 
superghosts were simply ignored, as usual in sigma-model constructions. 
That is also correct, because it amounts to ignoring an overall volume of the 
super-M\"{o}bius group, which is in fact finite \at. In the RR-vacuum 
considered in this section the ghosts must be considered; fortunately, 
we see that they introduce no significant complications.  

\subsec{\bf RR-couplings of the D-brane}
We now have a wavefunctional representation of the RR-vacuum.
The next step is to project the result \th\ on to a boundary wave
functional.
% which in general breaks conformal invariance, taking the theory 
%off-shell. 
As an example we begin by considering a single BPS D-brane,
including its world-volume gauge field. The boundary action is 
simply \ai\ 
%\eqn\bactbps{
%S_{\rm bndy}= -i\int d\tau (A_{\mu}{\dot x}^{\mu}
%+{\alpha^\prime\over 2}F_{\mu\nu}\psi^{\mu}\psi^{\nu})~,
%}
and the mode expansions are \d\ for the bosons and 
\eqn\modebps{
\psi^{\mu} = 
\sum_{n=-\infty}^{\infty}\psi_{n}^{\mu}e^{in\tau}~
}
for the periodic fermions. 
For constant field strength 
the boundary wave functional is therefore
\eqn\bpsd{
\Psi^{BPS}_{\rm bndy} = e^{-S_{\rm bndy}}=
\exp\left[-2\pi\alpha^{\prime}F_{\mu\nu}
[\sum_{n=1}^{\infty}({1\over 2}nX_{-n}^{\mu} X^{\nu}_{n}-
i\psi_{-n}^{\mu}\psi_{n}^{\nu})-
{i\over 2}\psi_{0}^{\mu}\psi^{\nu}_{0}]\right]~.
}
Except for the integer moding, the derivation of this result is 
identical to the NS-NS sector computation leading to the Born-Infeld action.

The projection of the bulk wave functional \th\ on to 
 \bpsd\ proceeds by integration over all field components.
The
integrals over non-zero modes are trivial in the present situation,
the contributions from bosons and the fermions cancelling by supersymmetry.
We are thus left with the zero-mode integrals
\eqn\zrrbps{
Z_{RR}^{BPS} = T_{D9} \int\! {\cal D}X_{0}{\cal D}\psi_{0}~
e^{2\pi\alpha^{\prime}{i\over 2}F_{\mu\nu}\psi_{0}^{\mu}\psi_{0}^{\nu}}
\sum_{p ~{\rm odd}}
{(-i)^{9-p\over 2}\over (p+1)!}
C_{\mu_{0}\cdots\mu_{p}}\psi^{\mu_{0}}_{0}\cdots\psi^{\mu_{p}}_{0}~.
}
The physically significant combination of the normalization ${\cal N}^{RR}$ 
and various measure factors was determined by comparison with the
known result for a single $D9$-brane without a gauge field. The bosonic 
zero-mode integral gives 
an overall volume integral and the remaining fermionic zero-mode integral is 
readily evaluated with the result
\eqn\zrrres{
Z_{RR}^{BPS} = T_{D9} \int\! C\wedge e^{2\pi\alpha^{\prime} iF}~,
}
in the familar representation as a formal sum of differential forms, 
{\it i.e}
\eqn\rrcdef{
C = \sum {(-i)^{9-p\over 2}\over (p+1)!}
C_{\mu_{0}\cdots\mu_{p}}dx^{\mu_{0}}_{0}\wedge\cdots\wedge dx^{\mu_{p}}_{0}
~.}
This is  the correct result, including coefficients. It was
previously derived using the boundary state formalism \bsrrrefs , 
a close relative to the present set-up, and independently by
anomaly inflow \ghm.

Suppose we allow $F_{\mu\nu}$ to be nonconstant and try to derive a 
generalization of \zrrres.  The new feature is that in \bpsd\ we replace
$F_{\mu\nu}$ by $F_{\mu\nu}(X)$. Bosons and fermions no longer cancel 
(because a generic zero-mode background $(X_0,\psi_0)$ breaks supersymmetry)
 and the integrals are non-Gaussian.   Nevertheless there is a
precise sense in which the formula \zrrres\ remains correct, but we defer 
discussion of this point to section 5.6.    

\subsec{\bf RR-couplings of the $D{\bar D}$-system}
We now compute the RR-couplings of the $D{\bar D}$ system.
 The wave functional describing  the bulk 
by definition does not depend on the boundary interactions
 so it is still \th. 
The boundary action describing the $D{\bar D}$ system is given in \azb:
\eqn\azbb{
S_{{\rm bndy}} = -\int\! d\tau \left[ {1\over 4}\dot{\eta}^{I}\eta^{I}
+\sum_{k=0}^{2m} {1\over 2k!}
(M_{1}-M_{0}^{2})^{I_1 \cdots I_k}\eta^{I_1} \cdots \eta^{I_k} \right]~.
}
Our task is to compute 
\eqn\qaa{Z_{RR}^{D\overline{D}} = {\cal N}_{RR}\int \! {\cal D}X {\cal D}
\psi {\cal D}\eta\,
e^{-S_{{\rm bndy}}}\,\Psi_{{\rm bulk}}^{RR}~,
}
with fields obeying periodic boundary conditions.  

What we now wish to 
argue is that it is justified to set all nonzero modes of 
$X^\mu$ and $\psi^\mu$
to zero.  
To proceed, allow a general $\tau$ periodicity, $\tau \sim \tau + \beta$, and
write \qaa\ in the canonical formalism as
\eqn\qb{Z_{RR}^{D\overline{D}} = {\rm Tr} (-)^F e^{-\beta H}~,
}
for some supersymmetric Hamiltonian $H$.  \qb\ is a Witten index \witind.
Because states of nonzero energy cancel between bosons and fermions,  
$Z_{RR}^{D\overline{D}}$ is independent of $\beta$, and is also constant
with respect to  smooth deformations of $H$ which preserve supersymmetry. 
Using our freedom to smoothly deform the theory, we will introduce
into \qaa\ a conventional looking $0+1$ dimensional kinetic term,
\eqn\qb{Z_{RR}^{D\overline{D}} = {\cal N}_{RR}\int \! {\cal D}X 
{\cal D}\psi {\cal D}\eta\,
e^{-S_0-S_{{\rm bndy}}}\,\Psi_{{\rm bulk}}^{RR},}
with
\eqn\qcc{S_0 = {1\over 4}\int \! d\tau \left((\dot{X}^\mu)^2 + \dot{\psi}^\mu 
\psi^\mu \right)  \sim 
\sum_{n=1}^\infty \left({n^2 \over \beta}X_{-n}^\mu X_{n}^\mu + 
in\psi^\mu_{-n}\psi^\mu_{n}
\right).}
Following a standard line of attack \index, we consider the path integral in the limit
$\beta \rightarrow 0$.  To avoid introducing spurious $\beta$ dependence,
the spacetime fields should be rescaled as one takes the limit.  Considering
the path integral for constant fields (in which case the path integral is
Gaussian) and demanding $\beta$ independence, one finds the rescalings:
$T^I \rightarrow \beta^{-{1 \over 2}}T^I$, 
$C_{\mu_{0} \cdots \mu_p}\rightarrow \beta^{{p+1 \over 2}}
C_{\mu_{0} \cdots \mu_p}.$  
Now, think of $S_0$ as supplying the propagators of the
theory, and the remainder as interaction terms.  Performing the 
$X^\mu$ and $\psi^\mu$ nonzero mode path integrals with no interaction
insertions gives $\beta^{-{d \over 2}}$, where $d$ is the spacetime 
dimension.  Next, note that we need to saturate $d$ zero modes integrals
of $\psi^\mu_0$.   From the form of the action and from the rescaling
of $C_{\mu_{0} \cdots \mu_p}$ one sees that each fermion zero mode 
is accompanied by a factor of $\beta^{1 \over 2}$, so saturating the
fermion zero modes precisely cancels the earlier factor of $\beta^{-{d \over 2}}$.
Now it is easy to see that $X^\mu$ and $\psi^\mu$ nonzero modes can be
dropped from the interaction terms,  since any of the associated diagrams
carry positive powers of $\beta$.   So we have arrived at the desired
result: in computing \qaa\ it is valid to set all nonzero modes of $X^\mu$
and $\psi^\mu$ to zero.

We have now reduced the computation to 
\eqn\qd{Z_{RR}^{D\overline{D}} = {\cal N}_{RR}\int \! 
{\cal D}X_0 {\cal D}\psi_0 {\cal D}\eta \,
e^{-S_{{\rm bndy}}} \sum_{p ~{\rm odd}}
{(-i)^{9-p\over 2}\over (p+1)!}
C_{\mu_{0}\cdots\mu_{p}}\psi^{\mu_{0}}_{0}\cdots\psi^{\mu_{p}}_{0}~,
}
with $S_{{\rm bndy}}$ given by restricting \azbb\ to zero-modes of $X^\mu$
and $\psi^\mu$.  Recalling the definition of $M$ in \ax\ we find that its
components, $M= M_0+\theta M_1$, are represented by the matrices 
\eqn\qe{M_0 = \sqrt{\alpha'}\pmatrix{iA^+_\mu \psi^\mu_0 & \overline{T} \cr
 && \cr
T & iA^-_\mu \psi^\mu_0} \equiv i\sqrt{\alpha'}{\cal A}~,}

\eqn\scdef{
M_1 = \alpha' \pmatrix{{i \over 2}(\partial_\mu A^+_\nu - \partial_\nu A^+_\mu)
\psi^\mu_0 \psi_0^\nu & \partial_\mu \overline{T} \psi^\mu_0 \cr
&& \cr
\partial_\mu {T} \psi^\mu_0 & {i \over 2}(\partial_\mu A^-_\nu - \partial_\nu 
A^-_\mu) \psi^\mu_0 \psi_0^\nu}\equiv i\alpha' d{\cal A}~.}
The matrix ${\cal A}$ is the superconnection \refs{\quillen,\book}.  
We will adopt  the notation of differential forms; e.g. for a $k$-form,
\eqn\qg{B^{(k)} = {1 \over k!} B_{\mu_{1} \cdots \mu_k}\psi_0^{\mu_1} \cdots
\psi_0^{\mu_k}~.}   Similarly, in \scdef\ $d$ denotes the exterior derivative.
Now note that the combination appearing in \azbb\ is
\eqn\qh{M_1 - M_0^2=
 i \alpha' (d{\cal A}- i {\cal A}\wedge {\cal A}) \equiv i\alpha' {\cal F}~.}
${\cal F}$ is the curvature of the superconnection,  given explicitly
by
\eqn\qi{ i{\cal F} = \pmatrix{iF^+ - T\overline{T} &
 D\overline{T} \cr DT & iF^- -\overline{T} T}~.}
Here $F^\pm$ are the full non-abelian field strengths,
\eqn\qj{ F^\pm = dA^\pm -i A^\pm \wedge A^\pm~,}
and the covariant derivatives are 
\eqn\qk{\eqalign{ DT &= dT + iTA^+ -i A^- T ~, \cr
 D\overline{T} &= d\overline{T} - i A^+ \overline{T} + i\overline{T}A^-~.}}
Here $TA^+$ and ${\bar T}A^-$ appear with an extra minus sign because,
as emphasized after \azb , we must multiply matrices using \wc . 
Now, upon performing the $\eta$ path integral,
 \qd\ becomes
\eqn\ql{Z_{RR}^{D\overline{D}} = 
{\cal N}_{RR}\int \! {\cal D}X_0 {\cal D}\psi_0  \, {\rm Str}~
e^{2\pi i \alpha' {\cal F}} \sum_{p ~{\rm odd}}
{(-i)^{9-p\over 2}\over (p+1)!}
C_{\mu_{0}\cdots\mu_{p}}\psi^{\mu_{0}}_{0}\cdots\psi^{\mu_{p}}_{0}~.
}
The supertrace arises because $\eta^I$ are periodic, and is defined by
\eqn\zh{ {\rm Str~}\!M = {\rm Tr}(-)^F M = {\rm Tr}\pmatrix{1 & 0 \cr 0 &-1}M.}
The remaining step is to do the integral over fermion zero-modes, which 
simply picks out the 10-form part of the integrand.  Hence our final 
result is
\eqn\kwform{Z_{RR}^{D\overline{D}} = T_{D9} \int \!  C \wedge{\rm Str}~
 e^{2\pi i \alpha' {\cal F}} ,}
where  factors of $i$ are included in the definition of $C$, as 
in \rrcdef .  
We fixed the overall normalization from the BPS computation.

The RR-couplings of the $D{\bar D}$-system were conjectured to be of the
form \kwform\ by Kennedy and Wilkins \kw . 
The evidence for the conjecture came from an S-matrix computation of the 
$d\overline{T} dT$ term.  Such an approach of course carries
with it some ambiguity when one attempts to write down an action which is
valid off-shell.  
  It is a welcome surprise that the full formula can 
be derived unambiguously rather simply from open string field theory. 
A  feature of the computation worth repeating is that we did not limit ourselves
to a linear tachyon profile nor to constant gauge field 
strength. This was possible because it was sufficient to
consider the $X^\mu$ and $\psi^\mu$ zero-modes. 
This in turn is closely connected to the fact that what we are computing
is the index of a certain operator, as we discuss in section 5.6.

The curvature of the superconnection satisfies some important properties
which make it suitable for appearing in the Chern-Simons term.  In
particular we note the Bianchi identity,
\eqn\qm{ {\cal D}{\cal F} = d{\cal F} -i {\cal A} \wedge {\cal F} + i
 {\cal F} \wedge {\cal A }=0~,}
and the transgression formula
\eqn\qn{ {\rm Str} {\cal F}^{n+1} = d\omega_{2n+1}({\cal A},{\cal F})~.}

As examples we consider some special cases of the RR-couplings.
If the tachyon vanishes the couplings become
\eqn\kwfor{
Z_{RR}^{D{\bar D}}(T=0) = T_{D9}\int\! 
C\wedge (e^{2\pi\alpha^{\prime}i F^{+}}-
e^{2\pi\alpha^{\prime}i F^{-}})~.
}
This is clearly recognized as the RR-couplings \zrrres\ of two 
BPS D-branes, one of each sign.

Next, take vanishing gauge fields and consider linear tachyon fields 
$\sqrt{\alpha^\prime}T^{1,2} = u_{1,2}x^{1,2}$. 
In section 4.1 we identified this configuration 
with a BPS $D7$-brane and we would like to check
that is has the correct charge. After expansion the partition sum 
\kwform\ becomes
\eqn\kwfr{
Z_{RR}^{D{\bar D}} = T_{D9}\int\! C\wedge 
e^{- 2\pi\alpha^\prime T{\bar T}} (2\pi\alpha^\prime)^2 
dT\wedge d{\bar T} = T_{D7}\int\! C_8~,
}
where $T_{D7}=(2\pi\sqrt{\alpha^\prime})^2 T_{D9}$ is  the correct
tension of a $D7$-brane. Note that the result is independent 
of the parameters $u_{1,2}$; in particular, the result is valid
before taking the fixed point limit $u_{1,2}\to\infty$.
This is a reflection of the fact that the D-brane charge
is a topological invariant.  

\subsec{\bf Example: the ABS Construction}
Consider the ABS construction \refs{\abs,\absapp} of 
a BPS $D(9-2m)$-brane on $2^{m-1}$ $D9-\bar{D9}$ pairs. The gauge 
fields vanish and the tachyons are
\eqn\abstach{
\sqrt{\alpha^\prime} \left(
\matrix{
	0 & {\bar T}  \cr
	T &   0
}\right) = u\gamma^i x^i~,
}
where $i=1,\ldots,2m$ is a vector index over the directions transverse to
the $D(9-2m)$-brane. The gamma matrices $\gamma^i$ represent the
Clifford algebra of the $SO(2m)$
transverse rotation group as $2^m\times 2^m$ matrices. They can be chosen
in the form
\eqn\gamdef{
\gamma^{i=1,\ldots,2m-1} = \left(
\matrix{
	0 & {\tilde\gamma}^i  \cr
	{\tilde\gamma}^i & 0 }\right) 
~,~~~~~
\gamma^{2m} = \left( \matrix{
	0 & -i I   \cr
	i I & 0 }\right)~,
}
where the $\tilde{\gamma}^i$ represent the $SO(2m-1)$ Clifford algebra.
The chirality matrix  ``$\gamma^5$''  is
\eqn\chidef{
\gamma^{2m+1} = \gamma^1\cdots\gamma^{2m} = 
 i^m  \left(
\matrix{
	I & 0  \cr
	0 & -I }\right)~.
}
The nonvanishing RR-couplings become
\eqn\absrr{\eqalign{
S &= T_{D9} \int \!C\wedge {1\over (2m)!} e^{-2\pi u^2 \vec{x}^2}
{\rm Str} \left(
2\pi\sqrt{\alpha^\prime}  u\gamma^i dx^i \right)^{2m} \cr
 &= T_{D(9-2m)}\int \!C_{10-2m}  ( u \int\! e^{-2\pi u^2 x^2}dx )^{2m}
i^{-m}{\rm Str} \gamma^{2m+1}  = T_{D(9-2m)} \int\! C_{10-2m}~,}
}
corresponding to a single $D(9-2m)$ brane, as expected. For $m=1$
the ABS construction reduces to the vortex solution considered
around \kwfr .

\subsec{\bf Non-BPS Branes}
The non-BPS D-branes are defined as projections by $(-1)^{F_L}$
of the $D{\bar D}$ system. Their RR-couplings are computed
as above, with the following substitutions:
\item{(1)} The tachyon is Hermitian,
$T={\bar T}$.  
\item{(2)}
The two gauge fields are identified, $A^+= A^- =A$.
% We take
%$F_{D{\bar D}}=F^+ + F^- = 2F_{{\rm non-BPS}~D}$.
\item{(3)}
After the above restrictions, the fermion $\eta^{2m}$ decouples.  One should
not perform the path integral over this fermion. 
%This corresponds to
%changing the overall normalization from $2T_{D9}$ to $\sqrt{2}T_{D9}$,
%which is the correct tension of a non-BPS D9-brane
\item{(4)}
The type IIA forms are defined as
\eqn\typac{
C = \sum_{\rm p~even} {(-i)^{8-p\over 2}\over (p+1)!}C_{\mu_0\cdots \mu_p} 
dx^{\mu_0}\wedge \cdots \wedge dx^{\mu_p}~.}
\noindent

%After elimination of the bosonic auxiliary fields the boundary action becomes
%\eqn\snbps{
%S_{\rm bndy} = 2\pi\alpha^\prime \left( T^{2} - {i\over 2}
%F_{\mu\nu}\psi^{\mu}_{0}\psi^{\nu}_{0}
%- D_{\mu}T\psi^{\mu}_{0}\eta \right)~,
%}
%where $T$ and $F_{\mu\nu}$ are $U(N)$ matrices. The covariant
%derivative is $DT = dT - i(AT+TA)$ or, recalling the convention
%that forms are taken to the left of group matrices, 
%\eqn\nbpscov{
%D_\mu T = \partial_\mu T - i[A_\mu, T]~. 
%}
%The boundary wave functional is now determined by carrying out
%the integral over the single $\eta$-fermion. After projection
%on to the bulk wave functional we find the couplings
Repeating the same steps as in the $D\overline{D}$ case, we find the
coupling
\eqn\npbsrr{
S = {T_{D9}\over\sqrt{2}} \int\! C\wedge {\rm Str} ~\exp \left[ 
2\pi\alpha^\prime \left(
\matrix{ i F - T^2 & DT \cr
 DT & i F- T^2}
\right) \right]~,
}
with the convention that the super-trace operation is defined 
with $\sigma_1$-type weight, {\it i.e.}
\eqn\stradrf{
{\rm Str}~M \equiv {\rm tr}~M
\left( \matrix{ 0 & I \cr I & 0 } \right)~,}
and the covariant derivative of the tachyon is now
\eqn\nbpscov{
D_\mu T = \partial_\mu T - i[A_\mu, T]~. }
It is clear that the result can be rewritten in terms of the
curvature of a superconnection of the form \scdef , with the
identifications detailed above.

In these considerations the origin of the non-BPS D-brane as a projection of
$D{\bar D}$ is emphasized. 
When interpreted as an object in its own right it is awkward to write 
the coupling of $N=2^m$ branes in terms of $2N\times 2N$ matrices. 
An alternative representation is obtained by writing the gamma matrices
as
\eqn\gamdefb{
\gamma^{i=1,\ldots,2m-1} = \left(
\matrix{
	 {\tilde\gamma}^i  &0 \cr
	0 &  {\tilde\gamma}^i  }\right).}
Then the  result becomes
\eqn\nbpsc{
S = \sqrt{2}T_{D9} \int\! C\wedge {\rm tr}~ 
e^{2\pi\alpha^\prime (i F - T^2 + DT)}~,
}
where the trace is an ordinary trace over group indices. Only odd 
forms are retained in the expansion of the exponential because
$C$ contains only odd forms in the type IIA theory. The linear
tachyon terms in \nbpsc\ were  discussed in \refs{\petr,\ben,\bsft}.

We end this section with a simple example, the $D8$-brane
as a domain wall on $D9-\bar{D9}$. Taking the tachyon 
$\sqrt{\alpha^\prime}T=ux$ and integrating over the linear profile 
we find
\eqn\deight{
S = \sqrt{2} T_{D9}
\int\! C_9 \wedge (e^{-2\pi\alpha^\prime T^2}~2\pi\alpha^\prime dT)
 = T_{D8} \int \!C_9~,
}
where $T_{D8} =  2\pi\sqrt{\alpha^\prime}T_{D9}$. The example verifies 
that the overall factors of $\sqrt{2}$ were incorporated correctly.

\subsec{\bf RR-charge as an Index}

Our computation of the RR-couplings boiled down to a path integral
in supersymmetric quantum mechanics with periodic boundary conditions. 
It is a famous result \index\ that various index theorems can be derived in
precisely this way.  In particular
\eqn\ra{ Z =\int\!{\cal D}\Phi \,e^{-S}= {\rm Tr}~(-)^F 
e^{-\beta H} = {\rm ind}(Q)~,}
where the index of the supercharge $Q$ counts the number of 
bosonic zero eigenvalues minus the number of fermionic zero eigenvalues.   
To prove an index theorem, on the one hand one writes $Q$ as an 
operator in the canonical formalism, and on the other hand evaluates 
the path integral expression for $Z$ in the $\beta \rightarrow 0$ limit.  

In section 5.3 we essentially computed the path integral in \ra .
The result was proportional to an infinite volume integral over $C^{(p+1)}$ 
which is factored out in the index computation. The remainder was 
the transverse integral over the generalized Chern character 
$\exp(2\pi\alpha^\prime i{\cal F})$. Previously we didn't pay close 
attention to the normalization of the zero-mode integral 
since this was absorbed into $Z_{RR}^{D\overline{D}}$; it is 
\eqn\ub{\left( { -i \over 4\pi^2 \alpha'}\right)^n~\int\! {\cal D}X_0 
{\cal D}\psi_0~,}
where we integrate over $2n=9-p$ transverse dimensions, 
$\mu = p+1 \cdots 9$. 

All that remains is to determine the operator form of the supercharge 
$Q$. The boundary action $S_0+S_{\rm bndy}$ in \qb\ has the detailed 
form
\eqn\rb{ S = {1 \over 4\alpha^\prime} 
\int\!d\tau d\theta D{\bf X}^\mu D^2 {\bf X}_\mu
 -\int\! d\tau \left[ {1\over 4}\dot{\eta}^{I}\eta^{I}
+\sum_{k=0}^{2m} {1\over 2k!}
(M_{1}-M_{0}^{2})^{I_1 \cdots I_k}\eta^{I_1} \cdots \eta^{I_k}\right]~,}
with $\mu$ running over the transverse directions. It is invariant 
under the supersymmetry transformations:
\eqn\rc{\eqalign{ \delta X^\mu &= \epsilon  
\sqrt{\alpha^\prime}\psi^\mu~, \cr
\delta \psi^\mu &= \epsilon {1\over\sqrt{\alpha^\prime}}\dot{X}^\mu~, \cr
\delta \eta^I &= \epsilon  F^I~,}}
where $F^I$ is expressed in terms of the $\eta^I$ as in \fondef .
%that we integrated out to arrive at \rb,
%\eqn\rd{F^I = -2 \sum {(-)^{k+1} \over 2 (k-1)!}
%M_0^{I_1 \cdots I_k} \eta^{I_2} \cdots \eta^{I_k}.}
Canonically quantizing \rb\ we find the commutations relations
\eqn\re{\eqalign{ [X^\mu,P_\nu] &= i \delta^\mu_\nu ~,\cr
\{\psi^\mu,\psi^\nu\} &= -2 \delta^{\mu\nu} ~,\cr
\{\eta^I,\eta^J\} &= 2 \delta^{IJ}~,}}
with the (Euclidean) momentum $P_\mu$,
\eqn\o{-iP_\mu = {1\over 2\alpha^\prime}\dot{X}_\mu -i  
\sum_{k~{\rm even}}{1 \over 2k!} A_\mu^{I_1 \cdots I_k}
\eta^{I_1} \cdots \eta^{I_k}~.} 
Now we want to show that the supercharge is
\eqn\q{Q = i\sqrt{\alpha^\prime}\psi^\mu P_\mu  - \sum {1 \over 2 k!}
M_0^{I_1 \cdots I_k} \eta^{I_1} \cdots \eta^{I_k}=
i\sqrt{\alpha^\prime}\psi^\mu P_\mu -i{\cal A} 
~.}
To show this we need to show that commuting fields with $\epsilon Q$ 
reproduces the variations in \rc. The first and third variations are 
easily derived. For the second recall that
\eqn\r{\eqalign{\sum {1 \over 2 k!}
M_0^{I_1 \cdots I_k} \eta^{I_1} \cdots \eta^{I_k}
&= i\sqrt{\alpha^\prime}\sum_{k~{\rm even}}{1 \over 2k!} \psi^\mu 
A_\mu^{I_1 \cdots I_k}\eta^{I_1} \cdots \eta^{I_k}
 \cr 
&\quad + \sqrt{\alpha^\prime}
\sum_{k~{\rm odd}}{1 \over 2k!} T^{I_1 \cdots I_k}\eta^{I_1} 
\cdots \eta^{I_k} .}}
%with $A^{I_1 \cdots I_k} = \psi^\mu A_\mu^{I_1 \cdots I_k}$.  
Then we see that
\eqn\s{[\epsilon Q, \psi^\mu] = -2i\epsilon \sqrt{\alpha^\prime}P_\mu + 
2i\epsilon \sqrt{\alpha^\prime}\sum_{k~{\rm even}}{1 \over 2k!} 
A_\mu^{I_1 \cdots I_k}\eta^{I_1} \cdots \eta^{I_k}=\epsilon 
{1\over\sqrt{\alpha^\prime}}\dot{X}^\mu~,}
as desired.
%Now, from general principles, it follows that the path integral for this 
%system computes ind(Q).  The index is defined as the number of bosonic
% zero eigenvalues minus the number of fermionic
% zero eigenvalues.  The fermion number operator is
% $\prod_\mu  \psi^\mu\prod_I \eta^I$.  
To write Q as a differential operator we use
\eqn\t{ P_\mu \rightarrow -i\partial_\mu, \quad
\psi^\mu \rightarrow i\gamma^\mu , \quad \eta^I \rightarrow \gamma^I,}
where $\gamma^\mu$ are spacetime gamma matrices, and $\gamma^I$ are
``internal'' gamma matrices.  Therefore, we can write Q as
\eqn\u{Q = \pmatrix{i\slash\!\!\!{\partial}+\slash\!\!\!\! A^+ & 
\overline{T} \cr T &
i\slash\!\!\!{\partial}+\slash\!\!\!\! A^- \cr}~.}
The index of $Q$ counts zero eigenvalues weighted by $(-)^F$. 
In contrast with \zh\ the $(-)^F$ here anti-commutes with all fermions 
and is given by $(-)^F=\prod_\mu\psi^\mu\prod_I \eta^I$. Choosing 
gamma matrices of the form \gamdef\ and using the rules \t\ we have
%
%Choosing a convenient off diagonal basis for gamma matrices, 
%the product over all gamma matrices is $\sigma_3 \otimes\sigma_3$ and
%so $(-)^F=\prod_\mu  \psi^\mu\prod_I \eta^I$  becomes
\eqn\v{(-)^F=\pmatrix{ 1 &&& \cr &-1&& \cr &&-1& \cr &&&1 \cr}~.}

We finally combine the ingredients. \ra\ was evaluated
first as a path integral with normalization \ub,
and then as the index of the expression \u\ for $Q$. We
arrive at the index theorem
\eqn\va{{\rm ind}\pmatrix{i\slash\!\!\!{\partial}+\slash\!\!\!\! A^+ & 
\overline{T} \cr T &
i\slash\!\!\!{\partial}+\slash\!\!\!\! A^- \cr}
= \left({-i \over 4\pi^2 \alpha'}\right)^{n}\int \! {\rm Str}~
e^{2\pi i\alpha'  {\cal F}}~.} 
Applying this result to tachyon condensation, we find that the net
D-brane charge of a solitonic configuration is equal to the 
index written  in \va.

Finally, we comment on the nature of corrections to  Chern-Simons 
couplings.  We nowhere assumed constant field strengths, so our formula
$\int\! C \wedge {\rm Str}~e^{2\pi i \alpha' {\cal F}}$ incorporates the
contribution of derivatives of open string fields.  However, it is important
to emphasize that this holds only for the integrated value of the coupling.
In other words, upon explicitly computing derivative corrections one
would arrive at a coupling (see for instance \wyllard )
\eqn\uc{ \int\! C \wedge \left\{{\rm Str}~e^{2\pi i \alpha' {\cal F}}
+ dV \right\}.}
This correction does not contribute to D-brane charges,
nor to open string correlation functions as long as $C$ is constant. 
So our results are exact for constant $C$ in this sense.  On the other
hand, if one were to make $C$ nonconstant then the correction would have
physical consequences.  

\newsec{Discussion}

We end this paper with a brief discussion of field redefinitions.
This issue is important if one
wants  to compare our results with those obtained
using different methods.  There are now  three approaches
to constructing D-branes as solitons in string field theory:
BSFT \bsft, level truncation \lumps, and noncommutative 
geometry~\refs{\hklm,\witncsft} --- all
presumably related by field redefinitions. 

For example, the tachyon potential
has been computed in level truncated superstring field theory
with the leading order result \berk\
\eqn\vlev{
V(T,{\bar T}) = 2T_{D9}~{1\over 2}(\alpha^\prime T{\bar T}-1)^2+
\cdots~.
}
Several of the corrections to this potential are known \corr\ and it is 
believed that the result to all orders is qualitatively similar.
An obvious difference is that in \vlev\ the minima lie at finite values
of $T$, whereas the minima in BSFT are at $|T| = \infty$. 
A more serious discrepancy is that level truncation violates gauge 
invariance since the gauge transformations act on the entire string
field, in contrast to BSFT and the effective field theories employed
in the noncommutative approach.   So it is clear that the field
redefinition relating the theories will have to involve all components
of the string field \david. 

The BSFT and noncommutative geometry approaches also realize gauge
invariance differently from each other, 
being related by some generalization of the
Seiberg-Witten map \sw.  In BSFT the soliton solutions have vanishing
gauge fields
%, which is possible because the tachyon derivative terms
%vanish in the closed string vacuum.  
so the RR-charge of solitons
arises entirely from the tachyon in this setup.  On the other hand,
a crucial element of soliton solutions in the noncommutative approach
at finite $B$ is the presence of a gauge field which sets all gauge
covariant derivatives to zero \hkl.  In this approach the gauge field
strength contributes to RR-charge.  
Apparently, the map between BSFT and noncomutative variables can
take a solution with vanishing gauge fields to one with non-vanishing
gauge fields. Formulating BSFT with a background $B$-field in terms
of noncommutative variables is a step in finding this field
redefinition \cori . 

The last issue we would like to address concerns 
tachyon derivative corrections to the action \senuniv\ 
\eqn\modbi{
S = \int\! V(T,{\bar T})\sqrt{{\rm det} (\delta_{\mu\nu}+F_{\mu\nu})}~.
}
It has been proposed that 
these can be accounted for by the substitution $F_{\mu\nu} \rightarrow
F_{\mu\nu} + \partial_\mu T \partial_\nu \bar{T}$ \genact .
This simple prescription does not 
occur in BSFT, but to settle the issue  one must 
determine whether a field redefinition exists that factorizes the action 
this way.

\bigskip\medskip\noindent 
{\bf Acknowledgements:}
We thank B. Craps, J. Harvey, E. Martinec, 
and particularly D. Kutasov for discussions.
After this work was completed the related papers \refs{\hori,\Tak}
appeared.      
This work was supported in part by NSF grant PHY-9901194 and by DOE grant
DE-FG0290ER-40560. FL was supported in part by a Robert R. 
McCormick fellowship. 

\listrefs

\end